\documentclass[preprint,journal]{vgtc}       % preprint (journal style)

\ifpdf%                                % if we use pdflatex
	\pdfoutput=1\relax                   % create PDFs from pdfLaTeX
	\usepackage{graphicx}                % allow us to embed graphics files
	\DeclareGraphicsExtensions{.pdf,.png,.jpg,.jpeg} % for pdflatex we expect .pdf, .png, or .jpg files
	\else%                                 % else we use pure latex
	\usepackage{graphicx}                % allow us to embed graphics files
	\DeclareGraphicsExtensions{.eps}     % for pure latex we expect eps files
\fi%

\graphicspath{{figures/}{pictures/}{images/}{./}} % where to search for the images

\usepackage{microtype}                 % use micro-typography (slightly more compact, better to read)
\PassOptionsToPackage{warn}{textcomp}  % to address font issues with \textrightarrow
\usepackage{textcomp}                  % use better special symbols
\usepackage{mathptmx}                  % use matching math font
\usepackage{times}                     % we use Times as the main font
\usepackage{cite}                      % needed to automatically sort the references
\usepackage{tabu}                      % only used for the table example
\usepackage{booktabs}                  % only used for the table example
\usepackage{wrapfig}
\usepackage{todonotes}
\usepackage{mathptmx}
\usepackage{graphicx}
\usepackage{times}
\usepackage{amsmath,amscd,amssymb, amsthm}
\usepackage{times}
\usepackage[abs]{overpic}
\usepackage{cases}
\usepackage{float}
\usepackage{caption}
\usepackage{subfloat}
\usepackage{subfig}
\usepackage{wrapfig}
\usepackage{placeins}
\usepackage{multirow}
\usepackage{paralist}

\onlineid{1685}

\vgtccategory{Research}
\vgtcpapertype{algorithm/technique}

\title{Automatic Polygon Layout for\\Primal-Dual Visualization of Hypergraphs}

\author{
Botong~Qu,
Eugene~Zhang, \textit{Senior Member,~IEEE,}
and Yue~Zhang, \textit{Member,~IEEE}}

\authorfooter{
\item
B. Qu is with the School of Electrical Engineering and Computer Science, Oregon State University. E-mail: qub@oregonstate.edu.
\item
E. Zhang is a Professor with the School of Electrical Engineering and Computer Science, Oregon State University. E-mail: zhange@eecs.oregonstate.edu.
\item
Y. Zhang is an Associate Professor with the School of Electrical Engineering and Computer Science, Oregon State University. E-mail: zhangyue@oregonstate.edu.
}

\shortauthortitle{Qu \MakeLowercase{\textit{et al.}}: Automatic Polygon Layout for Primal-Dual Visualization of Hypergraphs}

\abstract{
$N$-ary relationships, which relate $N$ entities where $N$ is not necessarily two, can be visually represented as polygons whose vertices are the entities of the relationships. Manually generating a high-quality layout using this representation is labor-intensive. In this paper, we provide an automatic polygon layout generation algorithm for the visualization of $N$-ary relationships. At the core of our algorithm is a set of objective functions motivated by a number of design principles that we have identified. These objective functions are then used in an optimization framework that we develop to achieve high-quality layouts. Recognizing the duality between entities and relationships in the data, we provide a second visualization in which the roles of entities and relationships in the original data are reversed. This can lead to additional insight about the data. Furthermore, we enhance our framework for a joint optimization on the primal layout (original data) and the dual layout (where the roles of entities and relationships are reversed). This allows users to inspect their data using two complementary views.
We apply our visualization approach to a number of datasets that include co-authorship data and social contact pattern data.
}


\CCScatlist{
\CCScat{K.6.1}{Management of Computing and Information Systems}%
{Project and People Management}{Life Cycle};
\CCScat{K.7.m}{The Computing Profession}{Miscellaneous}{Ethics}
}

\teaser{
\centering
\subfloat[][primal view]
{\includegraphics[height=1.6in]{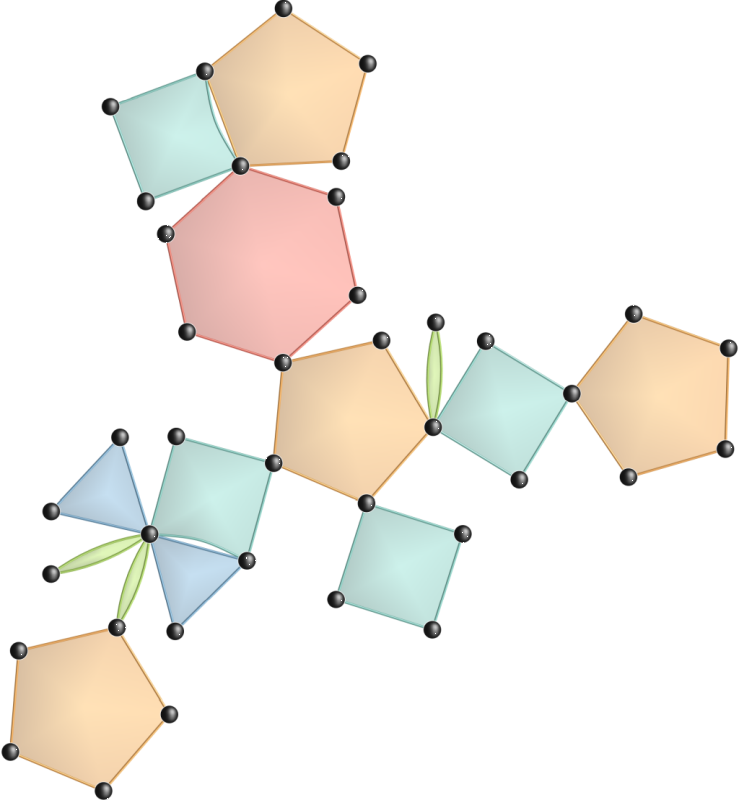}}\hspace{0.05in}
\subfloat[][dual view]
{\raisebox{0.1in}{\includegraphics[height=1.44in]{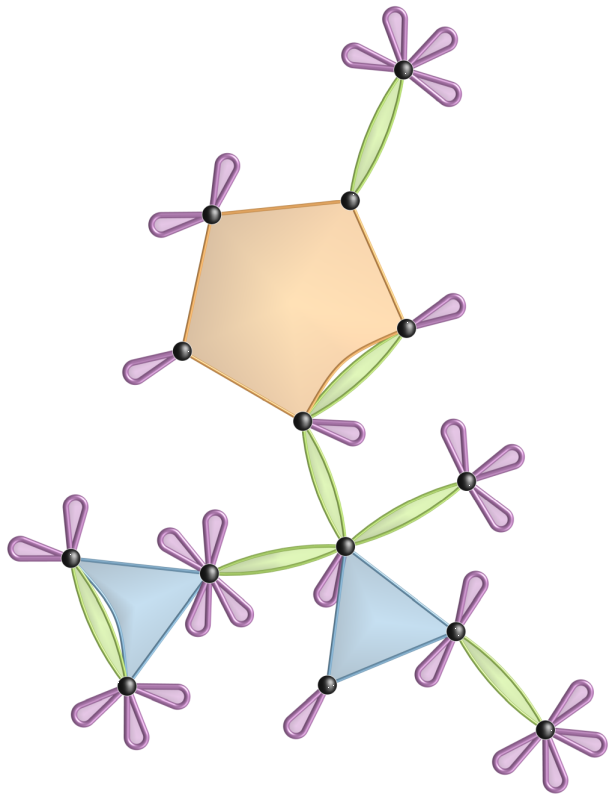}}}\hspace{0.05in}
\subfloat[][jointly optimized primal view]
{\raisebox{0.14in}{\includegraphics[width=0.25\linewidth]{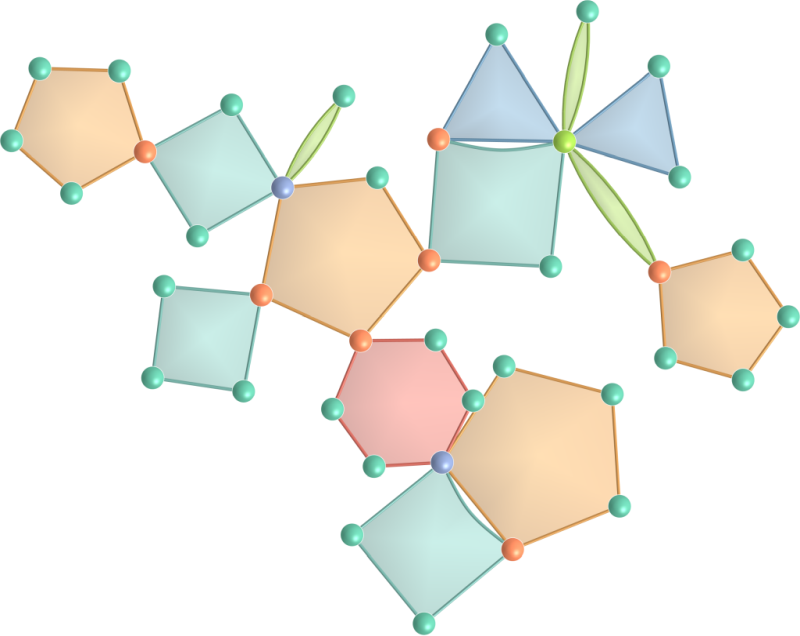}}}\hspace{0.05in}
\subfloat[][jointly optimized dual view]
{\raisebox{0.14in}{\includegraphics[width=0.25\linewidth]{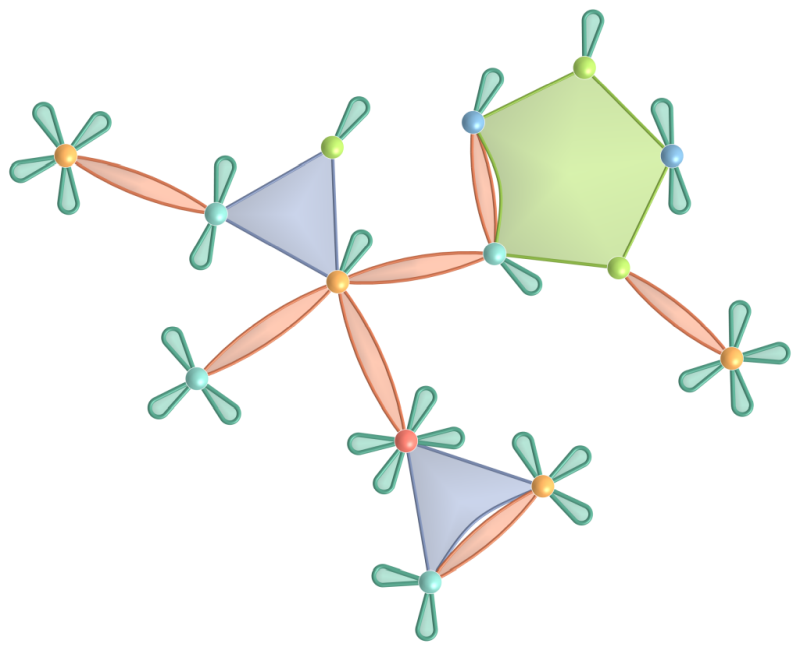}}}
\caption{A polyadic dataset with 39 entities and 14 relationships is visualized using a polygon-based approach~\cite{Qu:2017}. In the primal view (a), each polygon represents a relationship and the vertices in a polygon represent the entities in the corresponding relationship. The layout is generated by applying our automatic optimization algorithm (Section~\ref{sec:optimization}). In (b), we show the {\em dual view} of the same data, in which the vertices represent the relationships and the polygons represent the entities in the data. The layout is also generated using our optimization algorithm. In (c) and (d), we show the primal and dual views after applying our joint optimization (Section~\ref{sec:joint_optimization}). The color of a vertex in one view is the same as that of its corresponding polygon in the other view.
}
\label{fig:teaser}
}

\vgtcinsertpkg

\begin{document}

\firstsection{Introduction}\label{sec:introduction}


\maketitle

%% \section{Introduction} %for journal use above \firstsection{..} instead

Polyadic relationships are prevalent in datasets from social networks and biology. A polyadic relationship can involve $N$ entities in the data and is also referred to as an $N$-ary relationship. For example, in an academic publication dataset, a paper can be considered as an $N$-ary relationship that involves all of its authors.

A polyadic dataset can be theoretically modeled as a {\em hypergraph}, in which an entity is a vertex and an $N$-ary relationship is a hyperedge of $N$ vertices, each of which corresponds to one entity in the relationship. Given the high-dimensionality and the rich structures usually associated with hypergraph data, they often need to be inspected using a variety of visual metaphors with their respective strengths. Region-based techniques are a class of approaches that emphasize the distribution of vertices and hyperedges in the data based on additional criteria~\cite{Alsallakh:16}. In this approach, a vertex in the data is represented as a point and a hyperedge as a region that encloses all of its vertices. However, it can be difficult to visually extract basic information such as the number of vertices in a hyperedge and the set of hyperedges that are incident to a given vertex or share at least one vertex with a given hyperedge. Such information can be important for underlying applications.

One recently developed approach~\cite{Qu:2017} uses an $N$-sided polygon to represent an $N$-ary relationship (hyperedge) whose entities (vertices) are the vertices of the polygon, which makes it easier to see individual $N$-ary relationships and how they are related to each other (e.g., no overlapping vertices or multiple overlapping vertices). The vertices are initially placed using a force-directed graph layout method~\cite{hu2005efficient}. However, this can lead to concave polygons in the layout, which can make it difficult to see the number of vertices in such polygons. Qu et al.~\cite{Qu:2017} provide a set of layout editing operations such as moving a vertex, moving a polygon, scaling a polygon, and rotating a polygon. However, this manual postprocessing step can be labor-intensive, which greatly limits the potential application to larger datasets. As an example (shown in the accompanying video), manual design of a dataset with $81$ vertices and $22$ hyperedges required over seven minutes.

In this paper, we address this issue with an automatic framework for the generation of polygonal layouts for polyadic datasets. Our framework includes the following novel aspects compared to~\cite{Qu:2017}:
\begin{enumerate}
  \item A number of design principles for a polygonal layout.
  \item A set of objective functions motivated by the aforementioned design principles.
  \item The visualization for unary relationships.
  \item An operation called {\em starrization} that we develop to prevent the folding of a polygon during the optimization process.
  \item An operation called {\em pair swap} that allows the explicit control over the {\em order} of the vertices in a polygon.
\end{enumerate}

Our optimization algorithm allows larger datasets to be visualized at a higher quality without manual editing. For example, for the aforementioned dataset where manual editing took over seven minutes, our optimization framework took less than one second (see the accompanying video). Table~\ref{tbl:manual_and_opt} in Section~\ref{sec:manual_and_opt} (Appendix) provides the timing comparison statistics between the manual postprocessing~\cite{Qu:2017} and our optimization method for this dataset and two other datasets.

In addition, recognizing the duality between the entities and relationships in a dataset, we provide our users with both the primal visualization and the {\em dual visualization}, in which the roles of the entities and relationships are reversed. The primal and dual visualizations can be generated independently (Figure~\ref{fig:teaser} (a-b)) or through a joint optimization process that we have developed (Figure~\ref{fig:teaser} (c-d)). A vertex (or a polygon) can be selected in either the primal view or the dual view, and its corresponding polygon (or vertex) is highlighted in the other view. Since certain tasks can be performed more easily in one view and some related tasks may be performed better in the other view, our dual visualization allows the user to take advantage of the availability of both views simultaneously.

To evaluate the effectiveness of our approach and optimization framework, we conduct a user study and compare our visualization results to three recent region-based visualization techniques whose software are available: EulerView~\cite{simonetto2009fully}, HyperVis~\cite{Arafat:17}, and the Zykov representation~\cite{zykov1974hypergraphs, ouvrard2020hypergraphs}. The layouts of Zykov are generated by applying a force-directed method \cite{fruchterman1991graph} to the wheel associated graph~\cite{Arafat:17}. The user study indicates that our visualization technique outperforms these techniques in terms of both accuracy and speed. We also apply our visualization to datasets from co-authorship and social contact patterns.

\section{Related Work}
\label{sec:prev_work}

A hypergraph is an extension of a graph. Many graph drawing algorithms exist~\cite{Gibson:13}, and they usually represent the vertices and edges in the data with geometric objects. In the visual metaphor of Qu et al.~\cite{Qu:2017}, polygons are used to represent hyperedges in the data.

There has been much work in hypergraph visualization~\cite{Alsallakh:16,Vehlow2017graph}. Many existing techniques are based on Euler and Venn diagrams~\cite{Rogers:08,Stapleton:12,Micallef:14}, which focus on showing whether and by how much two hyperedges in the data intersect. For $N$-ary relationship visualization, it is important to show not only whether two relationships (hyperedges) share common entities (vertices), but also which entities are part of a given relationship.

Matrix-based techniques~\cite{Kim:2007,Sadana:2014,Lex:2014} model a hypergraph using a table. The hyperedges and/or vertices are mapped to a row, a column, or an entry in the matrix. When the number of rows and/or the number of columns is relatively large, it can be difficult to count the number of non-empty entries in a row or column or to decide whether two entries are in the same row or column. The effectiveness of this approach depends on whether the rows and columns can be sorted in a certain way~\cite{Alsallakh:16}. In addition, some matrix-based techniques such as the UpSet method~\cite{Lex:2014} do not explicitly model the vertices in the data. Instead, they are inferred from the intersections of hyperedges.

Another approach treats the hypergraph as a bipartite graph~\cite{Stasko:07,Dork:12,Alsallakh:2013}, in which both the vertices and the hyperedges are the nodes in the graph, though of different colors. Thus, the visualization of hypergraphs is converted to the visualization of bipartite graphs. Addressing the often large number of edge crossings in the visualization is the main challenge one faces using such an approach.

As an extension to the Euler and Venn diagrams, the subset-based approach~\cite{Santamara:2010,Riche:2011,Alsallakh:2013,Arafat:17} visualizes each hyperedge as a simple loop that defines a region. The vertices incident to a hyperedge are visualized as points enclosed by the loop. In some of the techniques~\cite{Riche:2011,Alsallakh:2013}, a vertex belonging to multiple hyperedges can be duplicated as multiple points. The copies of the same vertex are connected using curves. The subset-based approach is designed to handle hyperedges with overlaps~\cite{Alsallakh:16}. However, certain properties of individual hyperedges such as their cardinalities are not explicitly represented in some of these visualization methods~\cite{Santamara:2010,Riche:2011}. In our work, we reuse the two-dimensional CW-complex approach of Qu et al.~\cite{Qu:2017} in which each relationship (hyperedge) is mapped to a polygon whose number of edges encodes the cardinality of the hyperedge. To alleviate the intensive labor associated with manual editing during layout generation, we aim to provide an automatic layout algorithm.

A more recent approach models a hypergraph as a metro map (transit map) in which each hyperedge is modeled as a metro line with its vertices being the metro stations on the line. This approach allows the rich techniques of transit map generation~\cite{Wu:2020} to be applied to hypergraph visualization. Frank et al.~\cite{Frank:2021} investigate the theoretically minimum number of crossings between different metro lines (hyperedges) while Jacobsen et al.~\cite{Jacobsen:2021} provide practical optimization algorithms for the fast generation of high-quality metro maps for given hypergraphs. While this approach can make it easier to see the overlaps between different hyperedges, it can be challenging to quickly see the cardinality of a hyperedge, especially when its corresponding metro line partially overlaps with other metro lines.

Evans et al.~\cite{Evans:2019} propose to represent a hypergraph using a set of 3D polygons such that each polygon corresponds to a vertex in the hypergraph. They further investigate the theoretical feasibility of such a representation. However, the technique is not demonstrated on any data due to its theoretical nature, nor is an algorithm provided in generating a visualization based on this approach.

There has also been work on cluster visualization and community visualization~\cite{Fagnan:12,Dogrusoz:13,Vehlow2013fuzzy}. In these setups, the nodes in the data are partitioned so that each node belongs to exactly one cluster or a pre-dominant community~\cite{Dogrusoz:96} while having connections to other nodes in the data, including those in other clusters or pre-dominant communities. The visualization then focuses on placing each cluster or pre-dominant community in such a way that there is a clear spatial separation of the clusters or the communities~\cite{Fagnan:12}. The positions of the nodes in each cluster or pre-dominant community can be further improved through local operations~\cite{Dogrusoz:13}. To increase readability, glyphs are used to replace clusters and communities in order to provide a more abstract visualization of the relationships between clusters and communities~\cite{Dunne2013,Vehlow2013fuzzy,archambault2008grouseflocks}. In hypergraphs, each vertex can belong to multiple hyperedges. Spatially grouping the vertices based on their clusters and pre-dominant communities can have the unintended side effect of downplaying community memberships that are not chosen as the pre-dominant ones. In our work, we consider all hyperedges important and visualize them as such. A number of interactive tools are available~\cite{Sun:2016,Zhao:2018,Sun:2019} that support the exploration of bicluster data. As the data is visually represented using graphs, cluttering can occur as a result of excessive edge crossings~\cite{Alsallakh:16}.

\section{Data Representation}
\label{sec:data_representation}

A polyadic dataset consists of a set of entities $V$ and a multiset of relationships $R$. Each relationship $r\in R$ is a subset of $V$, and each entity must belong to at least one relationship. An entity can have attributes such as importance. Note that we allow two relationships to have an identical set of entities. We partition $R$ as $R=R_1 \cup R_2 \cup ... \cup R_{|V|}$ where $|V|$ is the number of elements in $V$. An element $r\in R_k$ ($k\ge 1$) is a subset of $V$ with $k$ elements. For example, $R_1$ consists of all unary relationships in the data and $R_2$ of all binary relationships. Note that we require a relationship to have at least one entity and at most $|V|$ entities. Furthermore, the order of the entities in an $N$-ary relationship is of little significance in our applications.

Polyadic data can be modeled as a hypergraph, with vertices (entities) and hyperedges (relationships). The polygonal representation of such data~\cite{Qu:2017} represents a hyperedge as a polygon, leading to a two-dimensional {\em CW-complex}~\cite{Hatcher:2002} for the hypergraph.
In the remainder of the paper, we will not differentiate between entities (semantics) and vertices (both theoretical and visual representations). Similarly, we will not differentiate among relationships (semantics), hyperedges (theoretical representation), and polygons (visual representation). More specifically, we will use vertices and polygons in the next sections.

The {\em degree} of a vertex is the number of polygons that contain this vertex. The {\em cardinality} of a polygon is the number of vertices in the polygon. Furthermore, the {\em degree} of a polygon is the number of polygons that share at least one vertex with the polygon. When a vertex is part of a polygon, they are {\em incident} to each other.

\section{Automatic CW-Complex Layout Optimization}
\label{sec:optimization}

We aim to provide an algorithm for the generation of a high-quality CW-complex layout for a given polyadic dataset. As with any optimization problem, we need the following components: (1) an objective function, and (2) an optimization framework. We describe both in this section.

\subsection{CW-Complex Layout Principles}
\label{sec:design_principles}

Before describing our objective functions, we state a number of design principles for CW-complex layouts that can produce clarity in the final layout.

\begin{enumerate}
\item Every polygon should be regular.~\label{principle_regular}
\item All polygons of the same cardinality should have the same area. ~\label{principle_uniform_area}
\item Polygons with larger cardinalities should have larger areas.~\label{principle_larger_area}
\item No polygon should contain self-overlaps and flips.~\label{principle_polygon_good}
\item Unnecessary overlaps between polygons should be avoided.~\label{principle_overcoverage}
\item When two polygons share at least three vertices, the intersection polygon should also be regular.~\label{principle_intersection_regular}
\item Two vertices should not overlap each other.~\label{principle_node_no_overlap}
\item A vertex should not appear on the border or in the interior of a polygon when the vertex is not part of the polygon. ~\label{principle_node_and_polygon_no_overlap}
\end{enumerate}

These principles are motivated by the following observations.

\begin{figure}[b]
  \centering
  \subfloat[][]{\includegraphics[height=1.0in]{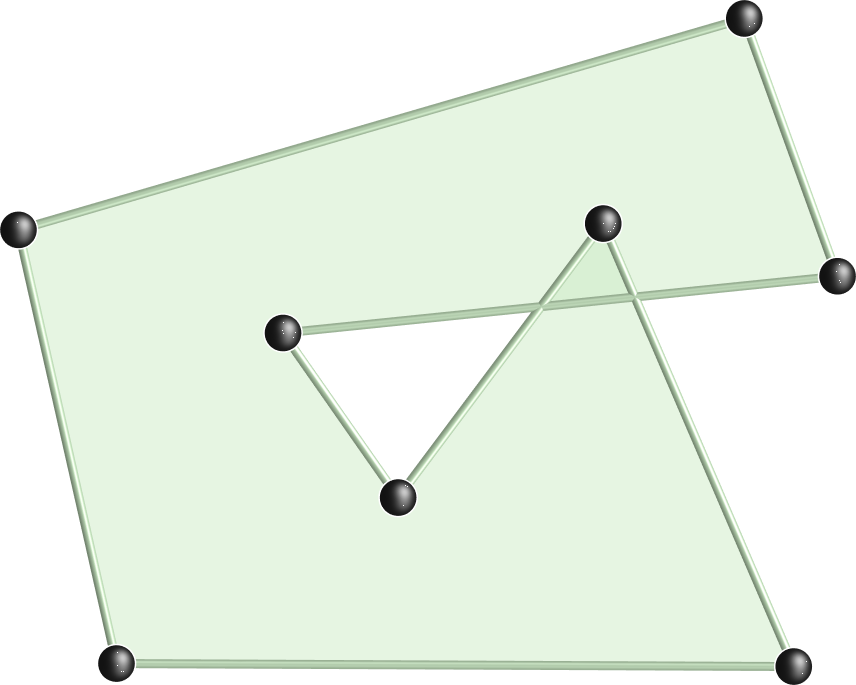}}\hspace{0.05in}
  \subfloat[][]{\includegraphics[height=1.0in]{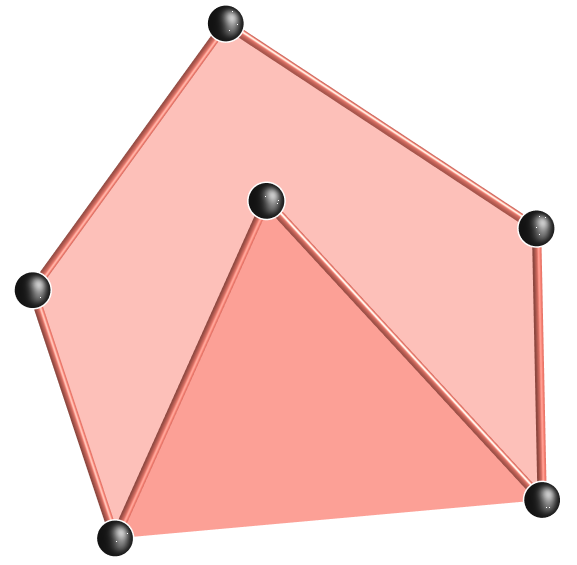}}
  \caption{Self-overlaps (a) and flips (b) in a polygon can make it difficult to recognize the cardinality of the polygon.   }\label{fig:overlap_problem}
\end{figure}

\begin{figure}[t]
     \centering
    $\begin{array}{@{\hspace{0.0in}}c@{\hspace{0.10in}}c@{\hspace{0.10in}}c}
     \includegraphics[width=1.0in]{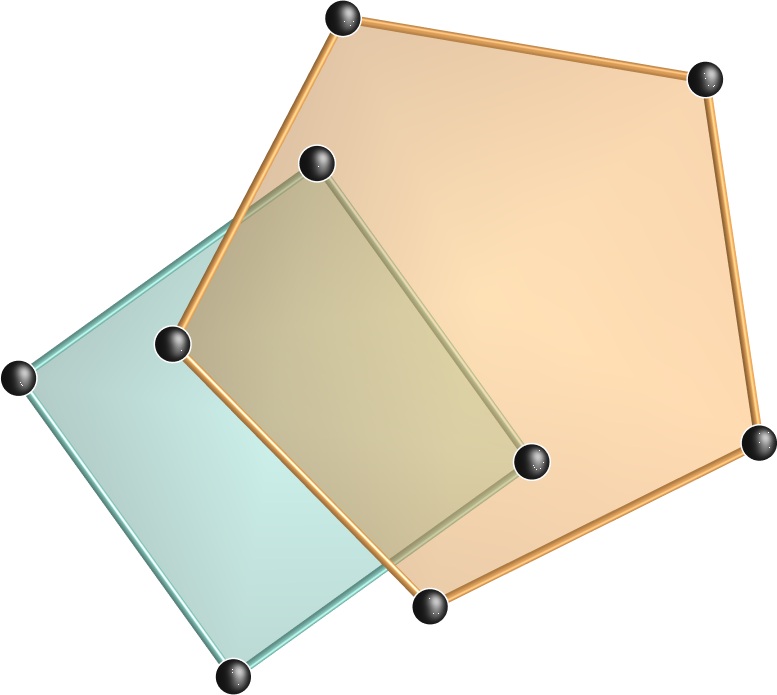}
    &\includegraphics[width=1.0in]{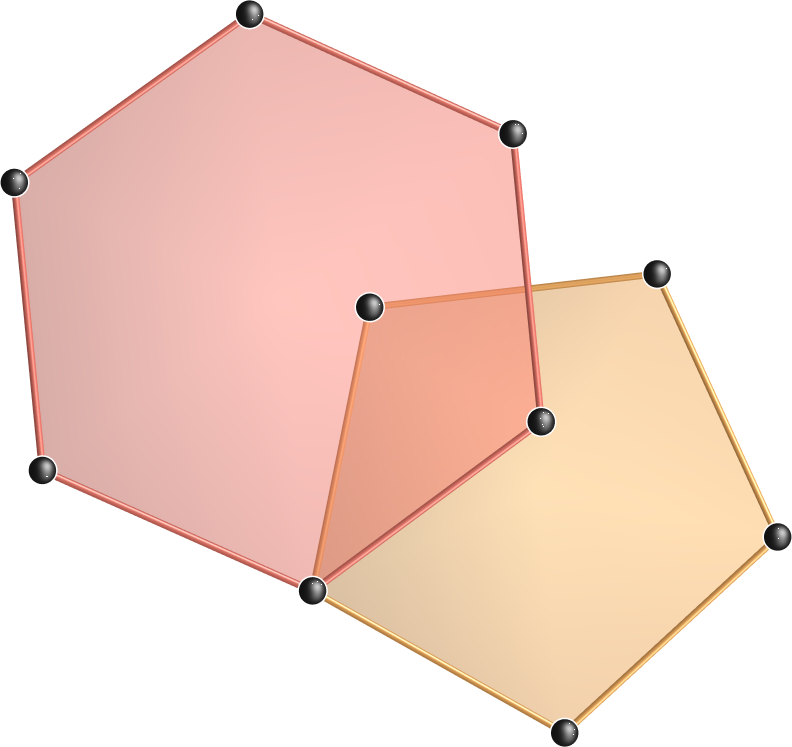}
    &\includegraphics[width=1.0in]{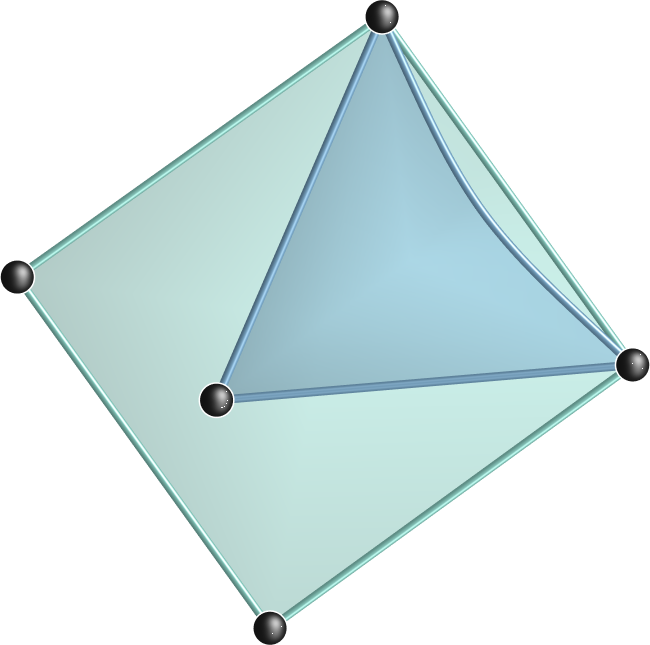}\\
    (a) & (b) & (c)\\
  \end{array}$
     \caption{This figure illustrates the motivations behind Principle~\ref{principle_overcoverage}. The two relationships share (a) zero, (b) one, and (c) two vertices, respectively.
          The interiors of the polygons partially overlap, which can distort the interpretation of the amount of sharing.
     }
\label{fig:overlappingcases}
\end{figure}

\begin{figure}[b]
  \centering
  \subfloat[][]{\includegraphics[width=0.3\linewidth]{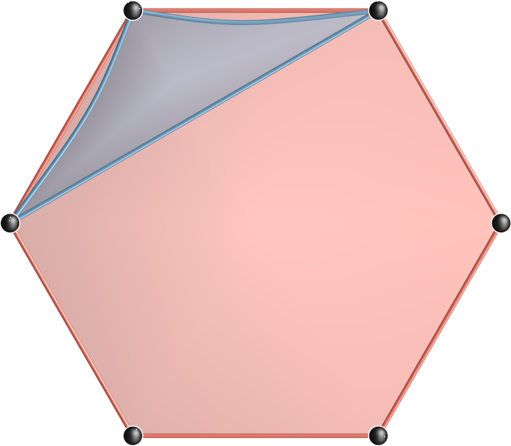}}\hspace{0.1in}
  \subfloat[][]{\includegraphics[width=0.3\linewidth]{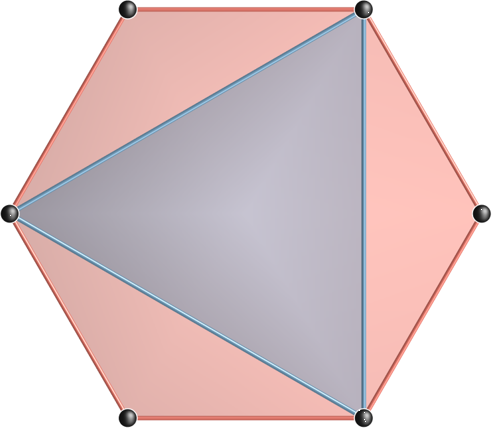}}
  \caption{When two convex polygons share at least three vertices, overlaps of the polygons are unavoidable. When the intersection polygon (shaded) is not regular (a), it can be more difficult to recognize the cardinality of the intersection polygon than when the intersection is regular (b).  }\label{fig:regular_intersection}
\end{figure}

First, the number of edges in a polygon can be best perceived when the polygon is regular (Principle~\ref{principle_regular}) and is sufficiently large. In addition to the number of sides in the polygon, we also wish to use the area of the polygon to encode the cardinality of the underlying $N$-ary relationship (Principles~\ref{principle_uniform_area} and~\ref{principle_larger_area}).

Second, a polygon with self-overlaps and flips (Figure~\ref{fig:overlap_problem}) can make it more difficult to recognize the cardinality of the polygon (Principle~\ref{principle_polygon_good}).

Third, when two polygons with at most two shared vertices overlap in the layout (Figure~\ref{fig:overlappingcases}), it can lead to the false impression that the polygons share more common vertices (Principle~\ref{principle_overcoverage}).

Fourth, when two polygons share at least three vertices and the polygon formed by the shared vertices is not regular, it can be more difficult to see the cardinality of the intersection polygon, i.e. the number of shared vertices (Principle~\ref{principle_intersection_regular}). Figure~\ref{fig:regular_intersection} contrasts two examples where the polygons' shared vertices form a regular polygon (b) and a highly non-regular one (a).

In addition, when two vertices overlap, they can appear as one (Principle~\ref{principle_node_no_overlap}). Similarly, if a vertex appears on the boundary or interior of a polygon not incident to it, the user can be given the false impression that the vertex is part of the polygon (Principle~\ref{principle_node_and_polygon_no_overlap}).

Principles~\ref{principle_larger_area},~\ref{principle_overcoverage}, and~\ref{principle_node_and_polygon_no_overlap} have been applied in existing subset-based hypergraph visualization, while the other principles that we have identified are rather specific to the polygon metaphor.

\subsection{Objective Function}

Based on the aforementioned design principles for hypergraph visualization, we formulate the following energy terms which are combined into the objective function used during the optimization.

\paragraph{Polygon Regularity (PR) Energy:} The {\em isoperimetric ratio} of a polygon~\cite{gage1984curve}, defined as $\frac{P^2}{A}$ where $P$ and $A$ are respectively the perimeter and area of the polygon, is a measure for the regularity of the polygon. Given an $n$-sided polygon, this measure is minimized when the polygon is regular (Principle~\ref{principle_regular}), which is $C_n = \frac{P^2}{A} = \frac{n^2}{\frac{n}{4}\cot(\pi/n)} =\frac{4n}{\cot{(\frac{\pi}{n}})}$. The isoperimetric ratio also favors convex polygons over non-convex polygons and over polygons with self-overlaps or flips.

Since the area of a polygon is a quadratic polynomial with respect to the $X$- and $Y$-coordinates of its vertices, the isoperimetric ratio is a radical (with square root terms). Using it directly as an energy can make the optimization process more challenging. Instead, we make use of the following $PR$ energy term:
\begin{equation}
E_{PR} = \sum_{\Gamma}P^2(\Gamma)-C_{|\Gamma|} \cdot A(\Gamma)
\label{eq:aspect_ratio}
\end{equation}
\noindent for all polygons $\Gamma$ in the data. $A(\Gamma)$, $P(\Gamma)$, and $|\Gamma|$ are the area, perimeter, and cardinality of $\Gamma$, respectively. This energy is non-negative and is zero only when $\Gamma$ is regular.

\paragraph{Polygon Area (PA) Energy:} To satisfy Principles~\ref{principle_uniform_area} and~\ref{principle_larger_area}, we observe that for regular polygons whose edge lengths are $1$, the area of the polygons is a monotonically increasing function of the cardinality of the polygons. That is, when combined with Principle~\ref{principle_regular} whose corresponding energy term is $E_{PR}$, we can formulate a $PA$ energy by requiring all edges in the layout to have the unit length as follows:
\begin{equation}
E_{PA} = \sum_{e}(l(e)-1)^2
\label{eq:aspect_ratio}
\end{equation}
where $e$ is any edge in any of the polygons in the data and $l(e)$ is the length of $e$. We will discuss how edges are generated in a polygon in Section~\ref{sec:initial_layout}. Note that while $E_{PA}$ itself does not ensure appropriate areas for polygons, it can do so when the polygon is regular. Therefore, combining the PR and PA energies can address Principles~\ref{principle_regular},~\ref{principle_uniform_area}, and~\ref{principle_larger_area}.

\paragraph{Polygon Separation (PS) Energy:} Principle~\ref{principle_overcoverage} suggests that when two relationships share at most two entities, it is desired to arrange the corresponding polygons in such a way to avoid having any intersection in the interior of the polygons. However, this is not always possible given the complexity of the data. To address this, we define the polygon separation (PS) energy for the case when the two relationships share zero, one, or two entities.

We approximate each polygon $\Gamma$ by a circle whose center is the centroid of $\Gamma$ and whose radius is the circumradius of a regular $n$-gon whose edge lengths are $1$. Note that in this case the circumradius is $\rho_n=\frac{1}{2}\csc\frac{\pi}{n}$.

\begin{figure}[tb]
  \centering
  \subfloat[][]{\includegraphics[width=0.30\linewidth]{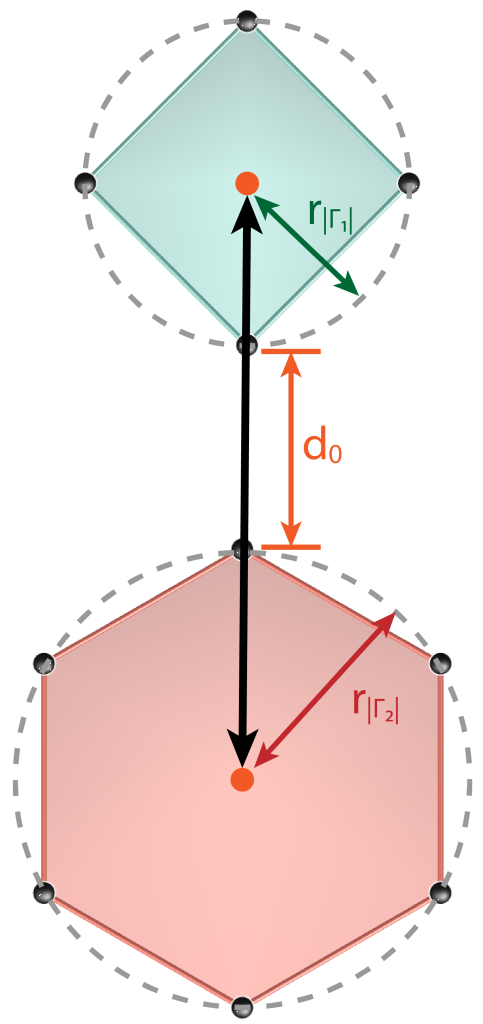}}\hspace{0.05in}
  \subfloat[][]{\includegraphics[width=0.30\linewidth]{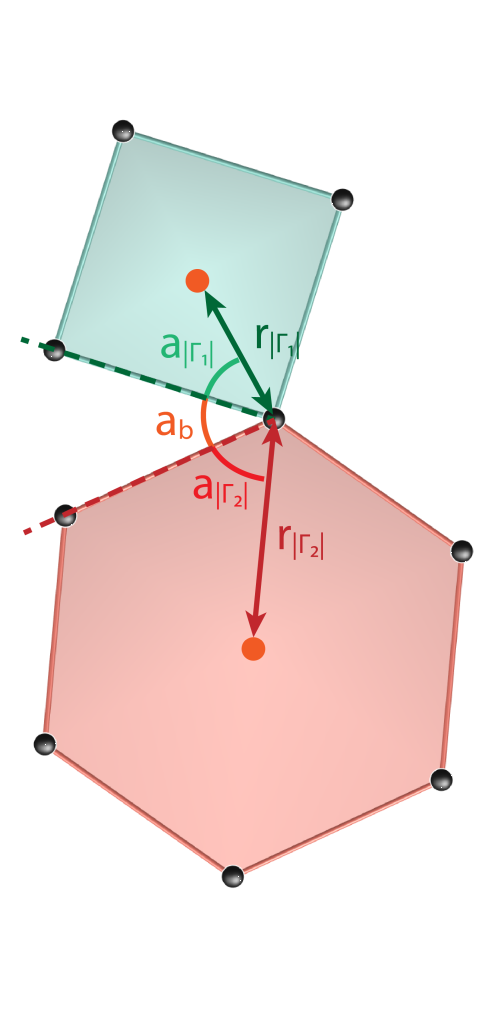}}\hspace{0.05in}
  \subfloat[][]{\includegraphics[width=0.30\linewidth]{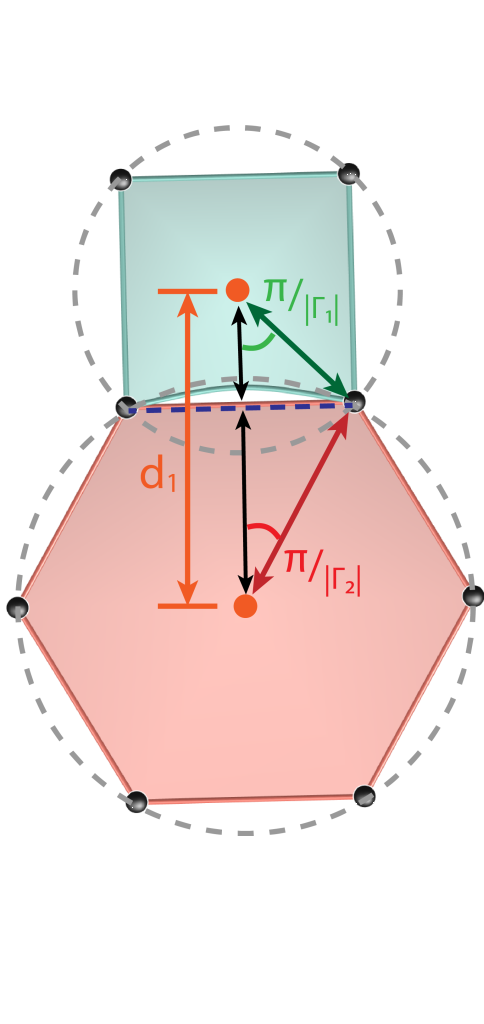}}
  \caption{The illustration of polygon separation energy when two polygons share zero (a), one (b), and two (c) vertices, respectively.
     }\label{fig:PS_illustration}
\end{figure}

When the two regular polygons $\Gamma_1$ and $\Gamma_2$ share zero vertices, as illustrated in Figure~\ref{fig:PS_illustration} (a), the minimal distance between their circumcenters is the sum of their circumradii. To prevent that the two polygons touch, we add a {\em buffer distance} $d_b$. Therefore, the minimal distance between the centroids of the polygons is $d_0(|\Gamma_1|, |\Gamma_2|)=\rho_{|\Gamma_1|}+\rho_{|\Gamma_2|}+d_b$, where $|\Gamma_1|$ and $|\Gamma_2|$ are the cardinalities of $\Gamma_1$ and $\Gamma_2$, respectively. Based on this analysis, we define the polygon separation energy for this pair of polygons as :
\begin{equation}\label{eq:ps_zero_overlap}
  E_{PS}(\Gamma_1, \Gamma_2)=f(d(\Gamma_1, \Gamma_2)- d_0(|\Gamma_1|, |\Gamma_2|))
\end{equation}
\noindent where $d(\Gamma_1, \Gamma_2)$ is the distance between the centroids of the polygons and $f(x)=\left\{ \begin{array}{ll} x^2 & \textrm{ if $x \le 0$} \\
0 & \textrm{ otherwise}
\end{array} \right .$.

When the two polygons share one vertex (the pivot), the distance between the polygons is measured in terms of the angle between the line segments formed by the pivot and the centroids of the two polygons (Figure~\ref{fig:PS_illustration} (b)). When the two polygons are regular, the minimal angle between the two line segments is $\pi\left(\frac{|\Gamma_1|-2}{2|\Gamma_1|}+\frac{|\Gamma_2|-2}{2|\Gamma_2|}\right)$. Similar to the previous case, we add a {\em buffer angle} $a_b$ such that the minimum angular distance between the two polygons is $a_0(|\Gamma_1|, |\Gamma_2|)=\pi(\frac{|\Gamma_1|-2}{2|\Gamma_1|}+\frac{|\Gamma_2|-2}{2|\Gamma_2|})+a_b$.

Let $a(\Gamma_1, \Gamma_2)$ be the angle between the two line segments in the current configuration. Then the polygon separation energy for this pair of polygons is
\begin{equation}\label{eq:ps_one_overlap}
  E_{PS}(\Gamma_1, \Gamma_2)=f(a(\Gamma_1, \Gamma_2)- a_0(|\Gamma_1|, |\Gamma_2|)).
\end{equation}

When the two polygons share two vertices (Figure~\ref{fig:PS_illustration} (c)), it is desirable that the shared vertices form an edge for both polygons. In this case, the ideal distance between the circumcenters of the polygons is $d_1(|\Gamma_1|, |\Gamma_2|)=\frac{1}{2}(\cot\frac{\pi}{|\Gamma_1|}+\cot\frac{\pi}{|\Gamma_2|})$. Therefore, we define the polygon separation energy in this case as
\begin{equation}\label{eq:ps_two_overlap}
  E_{PS}(\Gamma_1, \Gamma_2)=f(d(\Gamma_1, \Gamma_2)- d_1(|\Gamma_1|, |\Gamma_2|)).
\end{equation}

While it is possible that the two shared vertices do not form an edge in one of the polygons, in Section~\ref{sec:initial_layout} we introduce operations that can change the order of the vertices in the polygons that may lead to the desirable vertex orders for the polygons.

Finally, when two polygons share at least three vertices, their interiors inevitably overlap. In this case, we define the PS energy to be zero.

The total polygon separation energy is thus defined as
\begin{equation}\label{eq:ps_all}
  E_{PS}=\sum_{\Gamma_1, \Gamma_2} E_{PS}(\Gamma_1, \Gamma_2)
\end{equation}
\noindent for all pairs of polygons in the data. Note that besides Principle~\ref{principle_overcoverage}, our PS energy also aims to address Principles~\ref{principle_node_no_overlap} and~\ref{principle_node_and_polygon_no_overlap}. That is, when two polygons are properly positioned, their vertices cannot overlap, nor can a vertex of one polygon appear in the interior of another polygon.

The separation among monogons is different since a monogon shares at most one vertex with other polygons. Besides, in contrast to other polygons, the location of the incident vertex is not the only factor that decides the drawing of a monogon.

\begin{figure}[t]
  \centering
  \subfloat[][]{\includegraphics[width=0.35\linewidth]{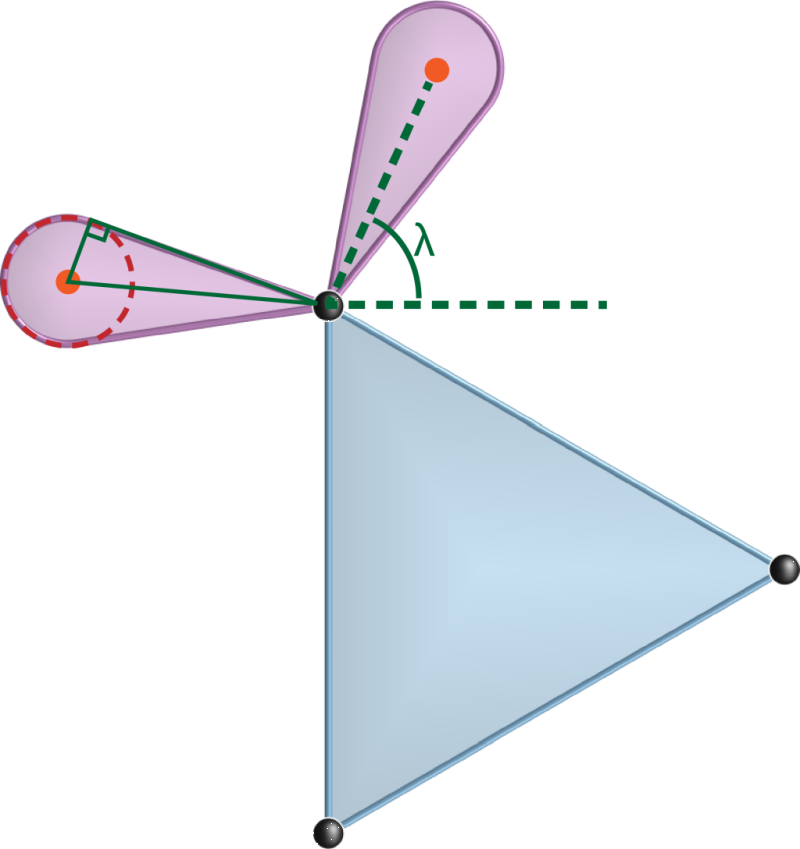}}\hspace{0.05in}
   \subfloat[][]{\includegraphics[width=0.35\linewidth]{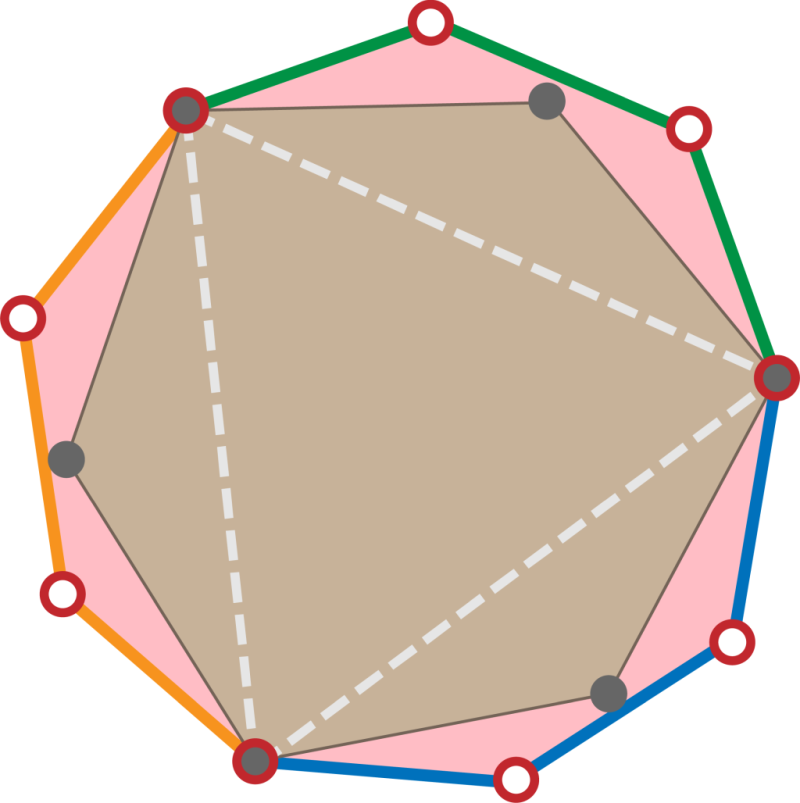}}
  \caption{(a) The orientations of monogons are optimized to decrease the repulsion between its center and other polygons' centers. The orientation of the monogon is measured by the angle $\lambda$ which is the angle between the $X$-axis and the line segment which connects the monogon center and the only incident vertex. The monogon center is the center of the semicircle tip of the waterdrop. (b) A hexagon (brown) and an enneagon (pink: nine-sided) share 3 vertices (filled circles with red boundary). The three shared vertices evenly separate the edges of the enneagon into three segments (colored differently).}
  \label{fig:monogon_PI_energy}
\end{figure}

Each monogon is drawn as the shape of a waterdrop which comprises a semicircle tip and two intersecting line segments tangent to the semicircle (Figure~\ref{fig:monogon_PI_energy} (a)). The intersection point of the line segments is the {\em vertex} of the monogon, while the center of the semicircle is the {\em center} of the monogon. In our system, all monogons have the same size, which is controlled by the radius of the semicircle in each monogon as well as the distance between the center and vertex of the monogon.

There are two degrees of freedom when drawing a monogon: the orientation angle $\lambda$ and its incident vertex's location. After optimizing the locations of vertices, one degree of freedom is fixed. However, a good orientation angle still needs to be decided for each monogon to reduce the overlapping between the monogon and its incident polygons.

We formulate the polygon separation energy for a monogon $\Gamma_1$ and a polygon $\Gamma_2$ as follows:
\begin{equation}\label{monogon_energy}
  E_{PS}(\Gamma_1, \Gamma_2) = \frac{w}{d(\Gamma_1, \Gamma_2)^2}
\end{equation}
\noindent where $w$ is a weight to control the separation force between different polygons. If both $\Gamma_1$ and its incident polygon $\Gamma_2$ are monogons, then $w$ is set to 0.1 to make monogons cluster together since we observe that this facilitates the counting of the monogons when needed. The weight $w$ is 1 otherwise.

This separation energy is calculated for every monogon and all of its incident polygons, so the total polygon separation energy for monogons is
\begin{equation}\label{monogon_energy}
  E_{PS(monogons)} = \sum_{\Gamma_1 \in R_1, \Gamma_1 \cap \Gamma_2\ne \emptyset}E_{PS}(\Gamma_1, \Gamma_2).
\end{equation}

\paragraph{Polygon Intersection (PI) Energy:} When two relationships share at least three entities, their polygons must overlap with an interior. Principle~\ref{principle_intersection_regular} states that the polygon formed by the shared vertices should be regular.

Let $\Gamma_0$ be the intersection polygon of $\Gamma_1$ and $\Gamma_2$. The vertices of $\Gamma_0$ divide the boundary of $\Gamma_1$ into $|\Gamma_0|$ segments, each of which is a collection of consecutive edges of $\Gamma_1$ (Figure~\ref{fig:monogon_PI_energy} (b)). For $\Gamma_0$ to stay regular, we need that each segment has the same number of edges of $\Gamma_1$. Clearly, this is only possible when the number of edges of $\Gamma_1$ is divisible by the number of edges of $\Gamma_0$. Still, we strive to distribute the edges of $\Gamma_1$ as evenly as possible into the above segments.

Let $s_i$ be the total length of segment $i$ where $1\le i \le |\Gamma_0|$. We define the {\em division energy} of $\Gamma_0$ with respect to $\Gamma_1$ as
\begin{equation}\label{eq:PI_subpolygon}
\mu(\Gamma_0; \Gamma_1)=\sum_{i=1}^{|\Gamma_0|} \left(s_i-\frac{|\Gamma_1|}{|\Gamma_0|}\right)^2
\end{equation}
\noindent where $\frac{|\Gamma_1|}{|\Gamma_0|}$ represents the ideal average length of each segment. This is under the assumptions that each edge in the polygon has a length of one and the perimeter of the polygon $\Gamma_1$ is $|\Gamma_1|$. We can similarly define the division energy of $\Gamma_0$ with respect to $\Gamma_2$, which we denote by $\mu(\Gamma_0; \Gamma_2)$.

The polygon intersection (PI) energy is then defined as
\begin{equation}\label{eq:PI_energy}
  E_{PI} = \sum_{\Gamma_1, \Gamma_2} \mu(\Gamma_1\cap \Gamma_2; \Gamma_1)+\mu(\Gamma_1\cap \Gamma_2; \Gamma_2)+E_{PR}(\Gamma_1\cap \Gamma_2)
\end{equation}
\noindent over all pairs of polygons $\Gamma_1$ and $\Gamma_2$ whose intersection polygon contains at least three vertices.

Our final objective function is thus $k_{PR} E_{PR} + k_{PA} E_{PA} + k_{PS} E_{PS} + k_{PI} E_{PI}$. To identify good values of the weights $k_{PR}$, $k_{PA}$, $k_{PS}$, and $k_{PI}$, we collect $19$ datasets. The number of hyperedges ranges from $23$ to $198$, and the number of entities ranges from $76$ to $527$. We sample $49$ weight sets $(k_{PR}$, $k_{PA}$, $k_{PS}$, $k_{PI})$ by bilinearly interpolating among four weight sets $(100, 1, 1, 1)$, $(1, 100, 1, 1)$, $(1, 1, 100, 1)$ and $(1, 1, 1, 100)$, which we consider the corners of a square. The square is then covered by a $7\times 7$ grid, each grid point represents a sampling weight. After conducting $49\times19$ experiments, we check which five weight sets could give us the average lowest sum energy without weighting. We get the final weight set by normalizing the average of these top five weight sets. Based on this experiment, we choose the following values: $k_{PR}=0.30$, $k_{PA}=0.16$, $k_{PS}=0.36$, $k_{PI}=0.18$.

Principle~\ref{principle_polygon_good} is automatically fulfilled since we do not allow self-overlaps and flips in any polygon. We describe how to achieve this in Section~\ref{sec:initial_layout}.

\subsection{Polygon Layout Optimization}
\label{sec:initial_layout}

Our optimization framework consists of the following stages. First, we generate an initial layout. Next, the layout is iteratively improved to achieve lower values with respect to the objective function. The process stops when some termination criteria are met. We now describe each of these steps in detail.

\paragraph{Initial Layout and Starrization:} Our system can generate different initial layouts, such as placing all the vertices in the data on a circle (circular initial layout) or randomly (random initial layout). We can also convert the hypergraph data into a graph by treating each hyperedge as a clique in the graph. The vertices are then placed by using a force-directed algorithm~\cite{hu2005efficient}.

Once all the vertices have been positioned, we need to construct the polygon for each relationship in the data. Note that the order of the vertices in the polygon is not meaningful with respect to the underlying relationship. However, randomly selecting the order of entities in a relationship can lead to a polygon with self-overlap or flip (Figure~\ref{fig:overlap_problem}), thus violating Principle~\ref{principle_polygon_good}.

\begin{figure}[t]
  \centering
%  \subfloat[][]{\includegraphics[height=1.1in]{images2021/glyphs/A082_SelfOverlap.png}}\hspace{0.05in}
%  \subfloat[][]{\includegraphics[height=1.1in]{images2021/glyphs/A023_self_intersection2_starrized.png}}
\subfloat[][]{\includegraphics[height=1.0in]{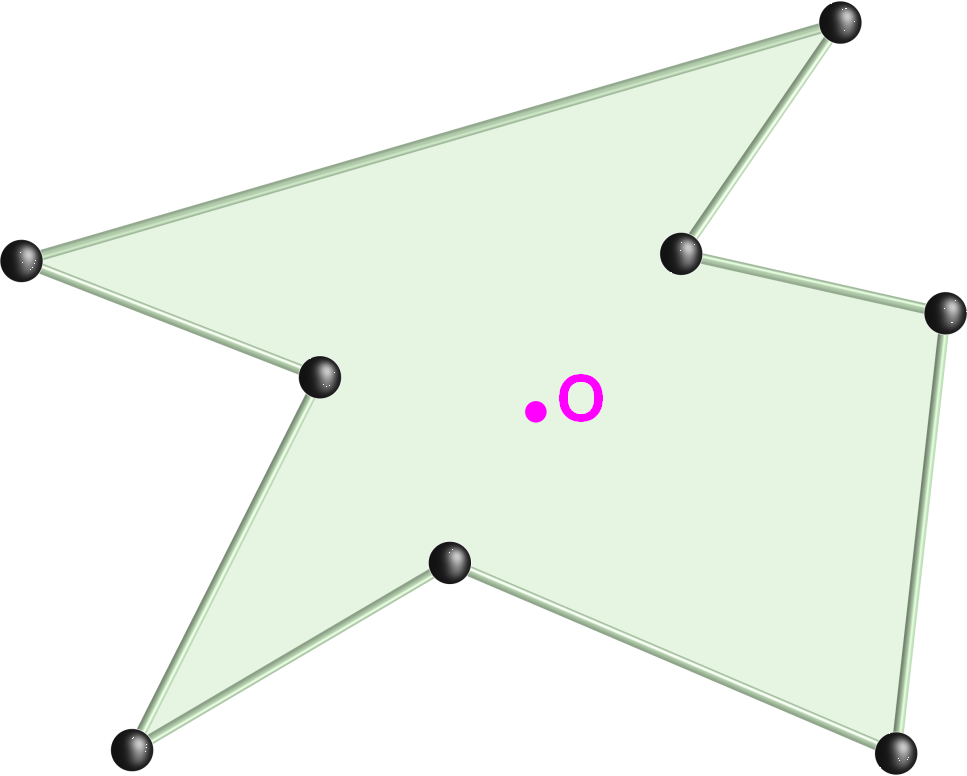}}\hspace{0.05in}
\subfloat[][]{\includegraphics[height=1.0in]{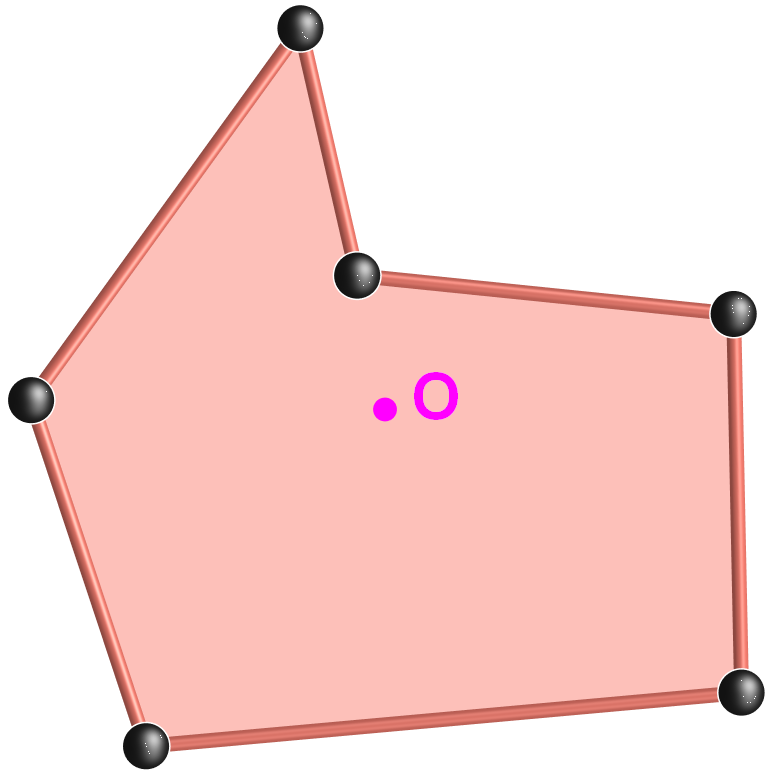}}
  \caption{The starrization operation changes the order of vertices of the polygon to eliminate self-overlaps and flips. Compare the examples in this figure to those shown in Figure~\ref{fig:overlap_problem}. The centroids, O, required by the starrization operation, are shown in both examples. }\label{fig:starrization}
\end{figure}

To address this issue, we employ a procedure which we refer to as the {\em starrization}. The input to this procedure is the positions of the vertices in a polygon without the connectivity information. The starrization then computes the centroid of the convex hull for the vertices of the polygon (Figure~\ref{fig:starrization}). The centroid is then used as a reference point, with the vertices in the polygon sorted by their angular coordinates. This gives rise to an order of the vertices in the polygon that is guaranteed to be free of self-overlaps and flips (compare the polygons before starrization in Figure~\ref{fig:overlap_problem} and the corresponding polygons after starrization in Figure~\ref{fig:starrization}). Note that our problem is related to but different from the problem of creating a polygonal region based on a set of points inside the region and a set of points outside the region~\cite{Reinbacher:2005}.

\paragraph{Energy Minimization:} Given the initial layout, our automatic layout algorithm iteratively improves the layout by finding new locations of the vertices in the data and the order of the vertices in the polygons. We treat the locations of the vertices as a variable vector in a $2|V|$-dimensional space where $|V|$ is the number of entities (vertices) in the data. The 2D coordinates of the vertices are consecutively encoded into the variable vector.

There are two operations performed. The first operation is to find a layout that yields the local minimum of energy. Because our objective functions are all arithmetic, we use the automatic differentiation library Adept\cite{hogan2014fast} to monitor the evaluation of the objective functions so that the gradient can be calculated automatically. With the capability to calculate the gradient, we choose to use the quasi-Newton optimization method L-BFGS \cite{liu1989limited} due to the speed and memory efficiency of this solver. During the optimization, the L-BFGS algorithm uses the gradient information to update the search direction and performs a one-dimensional search for the minimum of the energy on the line in the search direction~\cite{more1994line}. The solver has been shown to be able to handle non-linear optimization problems very well\cite{liu2009centroidal}. Note that starrization is performed when evaluating energies for potential new locations of vertices. This ensures that no self-overlaps and flips can occur after the optimization.

The second operation evaluates, for each pair of vertices in the same polygon, the objective function before and after the two vertices' locations are swapped. Note that the orders of the vertices in the polygon are also swapped. While a pair swap does not impact the shape of the polygon, it can impact other polygons that are incident to the vertices. Consequently, we perform starrization to the adjacent polygons to ensure no self-overlaps and flips after the swap. If a swap improves the objective function, the swap is accepted. Otherwise, it is rejected. Note that each pair of vertices inside each polygon is evaluated until no swap improves the energy.

The optimization alternates between the iterative line searches and the pair swaps. This process continues until there is no further gain in the objective function by the line searches and the pair swaps. Note that since our optimization problem is multi-objective and non-convex, there is no guarantee that a global minimum can be found. %\revised{W}e \revised{ have} observe\revised{d, however,} that our procedure converge\revised{s} \revised{with}in \revised{a} reasonable amount of time.

Note that translating or rotating a layout does not change its evaluation with respect to the objective function. Consequently, we fix the locations of two vertices while optimizing for the positions of the other vertices during a line search.

Our default initial layout is based on the force-directed algorithm of Hu\cite{hu2005efficient}. However, while different initial layouts can lead to different optimized layouts, it is not clear which initial layout consistently outperforms the other ones. Consequently, we provide different initial layout options in our system. Figures~\ref{fig:initial_layout_FD},~\ref{fig:initial_layout_circle}, and ~\ref{fig:initial_layout_random} in Section~\ref{sec:appendix} (Appendix) compare various optimized layouts (right) based on the three initial layout schemes (left). Given that the problem of finding a polygonal layout with minimal polygon overlaps is similar to the crossing number problem\cite{Garey:83}, which is known to be NP-hard, it is unlikely that any of the initial layouts guarantees to generate the global optimal layout. We observe that the force-directed scheme tends to lead to fewer overlaps in the polygons. On the other hand, the circular and random schemes tend to generate layouts that are space-filling. There is a trade-off between space utilization and the amount of unnecessary overlaps. We have made all three options available in our system, and used the force-directed method to generate the layouts for our user study.

\section{Dual-Views and Joint Optimization}
\label{sec:joint_optimization}

While the polyadic data pre-determines the roles of the entities and relationships, such a role assignment can be arbitrary. For example, in a paper--author dataset, a reader may consider each author as an entity and each paper as a relationship over its authors. On the other hand, a researcher might consider his/her papers as the entities while the authors as relationships over the papers they have authored. The paper-centric view and the author-centric view can be seen as ``the two sides of the same coin''. In general, any data can have two views: (1) the primal view based on the input data model, and (2) the dual view in which the roles of the entities and relationships in the data are reversed.

Given these observations, we provide a synchronized optimization and visualization framework in which both viewpoints are not only optimized and displayed individually (e.g. Figure~\ref{fig:teaser} (a) and (b)), but also optimized jointly and displayed side-by-side (e.g. Figure~\ref{fig:teaser} (c-d)).

When generating the layout for the dual view, we wish to position each polygon's dual element, which is a vertex in the dual view, as close as possible to this polygon's centroid in the primal view. This allows the correspondence between the polygon and its dual vertex to be relatively easily perceived. To achieve this goal, we define the following energy.

\paragraph{Dual Distance Energy (DD).} Let $Dual(v)$ be the polygon dual to $v$. For a polygon $\Gamma$, let $O(\Gamma)$ be its centroid.
The dual distance (DD) energy for the whole layout is then defined as
\begin{equation}\label{eq:dual_distance}
  E_{DD} = \sum_{v_i \in V} (O(Dual(v_i)) - v_i)^2.
\end{equation}
\vspace{-0.2in}
\noindent

With the dual distance energy, the objective function becomes $k_{PR} E_{PR} + k_{PA} E_{PA} + k_{PS} E_{PS} + k_{PI} E_{PI} + k_{DD} E_{DD}$ where $k_{PR}=k_{PA}=k_{PS}=k_{PI}=k_{DD} = 0.2$ by applying equal weights to scalarize the multi-objective function into one scalar function. Note that our system allows users to change the weights to prioritize different design principles during the optimization.

When conducting the joint optimization, we optimize the two views simultaneously. That is, the locations of the vertices in the primal view and the locations of the vertices in the dual view are formed as one high-dimensional vector and used to evaluate the objective function that includes all the energy terms for the primal view and the dual view as well as the DD energy. A new configuration (a layout for the primal and a layout for the dual) is accepted only if the total energy decreases. This applies to both the line search and the pair swap operation.

In both the primal and dual views, a monogon corresponds to a degree-one vertex in the other view. To assist the mapping between monogons and their dual elements, the objective function $E_{DD}$ (Equation~\ref{eq:dual_distance}) can be applied again locally.

We still use the same optimization solver to minimize the energy for monogons. However, for monogons, the input becomes a vector of $|R_1|$ variables where $|R_1|$ is the number of monogons in the hypergraph. Each variable corresponds to the orientation angle $\lambda$ of a monogon. Note that when optimizing the monogons, all other terms of our energy are no longer useful since a monogon is always regular, in the ideal length, and shares only one vertex with any other polygon.

\section{Visualization and Functionalities}
\label{sec:visualization}

Our visualization system allows the user to load a hypergraph and visualize it. To increase perceptibility, each vertex is rendered as a sphere with reflective material. Each edge is rendered as a cylinder that is also reflective. Each polygon is rendered with reflective and translucent material. This leads to rendering effects that are similar to the Cushioned Treemaps~\cite{vanWijk:1999}.

When multiple polygons intersect, polygons with larger cardinalities are behind those with smaller cardinalities. This choice is similar to that of Kelp Diagrams~\cite{Dinkla:2012}.

The colors of the polygons can be based on the properties of the underlying relationships. To increase the readability of overlapping polygons, we use colors suggested by ColorBrewer~\cite{harrower2003colorbrewer}.

We provide two views, one for the primal representation and the other for the dual representation. The user can choose to see either view only or both. When both views are shown, the user can also select a vertex or a polygon in one view to inspect its properties. The corresponding polygon or vertex in the other view is highlighted automatically.

\begin{figure}[t]
  \centering
    \subfloat[][Q1]{\includegraphics[height=1.5in]{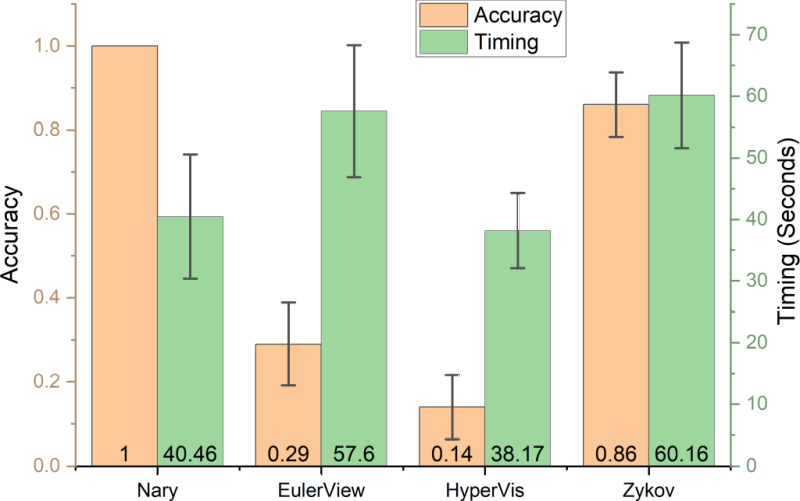}}\hspace{0.02in}\\
	\subfloat[][Q2]{\includegraphics[height=1.2in]{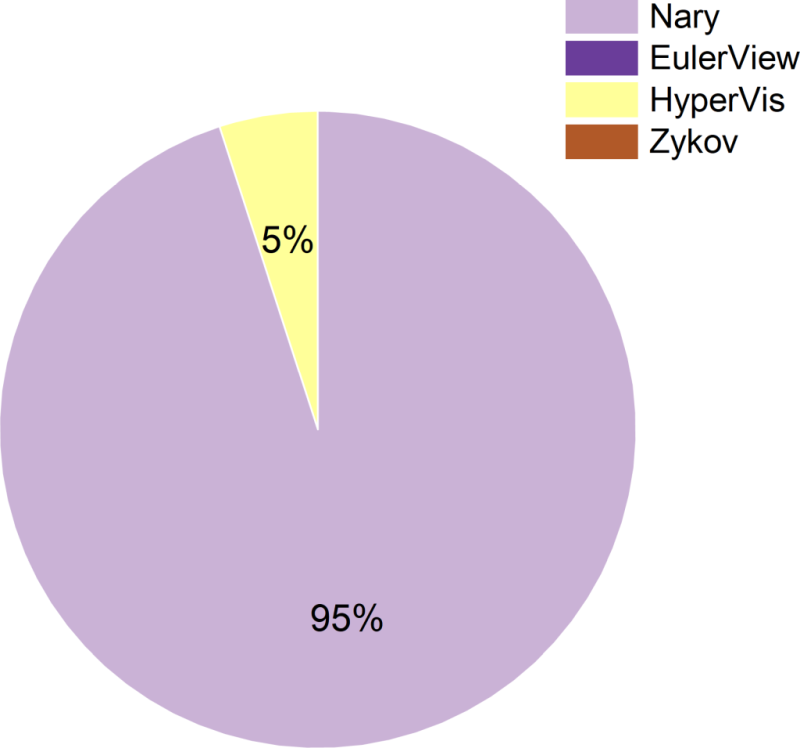}}\hspace{0.02in}
	\subfloat[][Q3]{\includegraphics[height=1.2in]{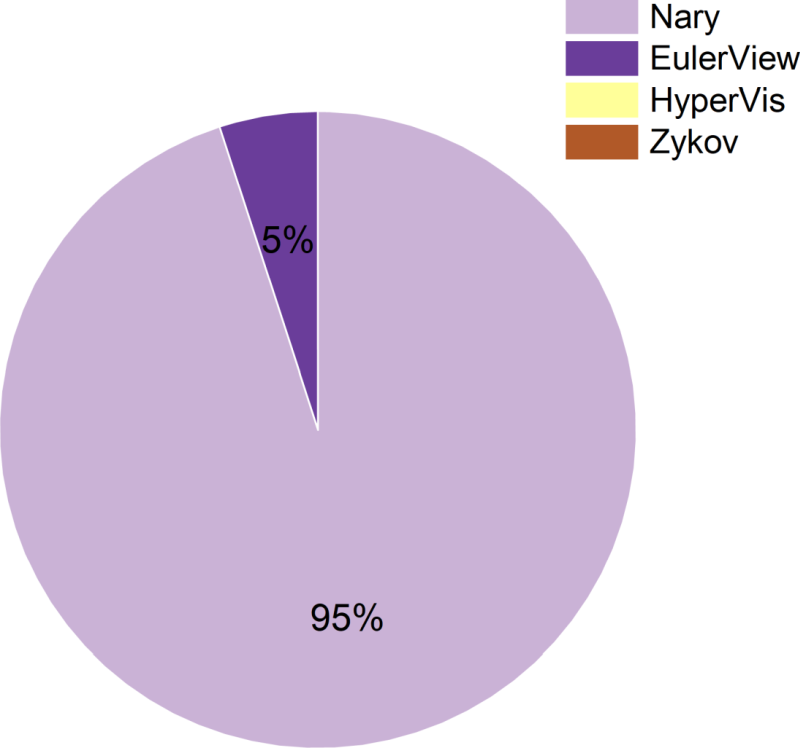}}\hspace{0.02in} \\
    \subfloat[][Q4]{\includegraphics[height=1.5in]{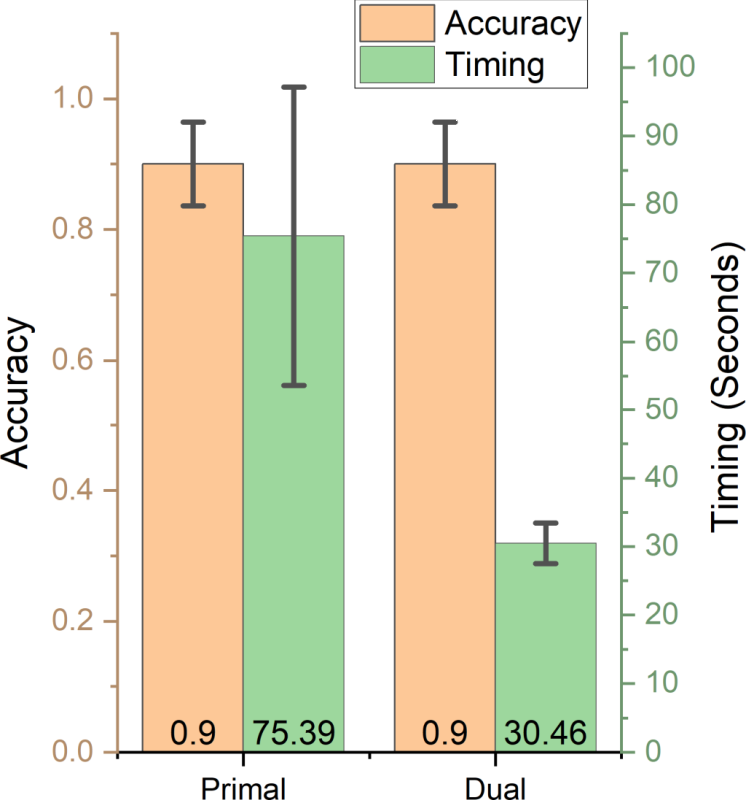}}\hspace{0.02in}
    \subfloat[][Q5]{\includegraphics[height=1.5in]{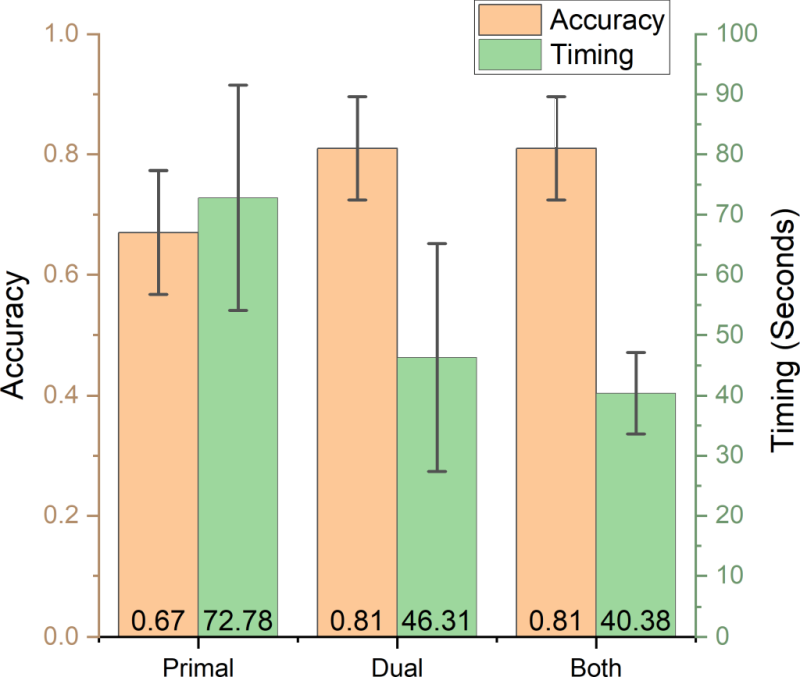}}
  \caption{This figure shows the resulting statistics of our survey with $21$ participants. For Q1, Q4, and Q5, the yellow bars represent the accuracy rate for each question and each method. The green bars represent the timing results (in seconds). The scale of accuracy is always shown on the left of the figure, and the scale of the timing is shown on the right. The standard error of the accuracy and timing results (yellow and green bars) are included as black line segments. Note that the standard error is zero for the polygon-based method in Q1. Consequently, the standard error in this case is not shown by the graphing software. The user preferences over the four methods inquired in Q2 and Q3 are  shown in pie charts.
}\label{fig:survey_results}
\end{figure}

\begin{figure*}[ht]
  \centering
    \subfloat[][our method]{\includegraphics[height=1.0in]{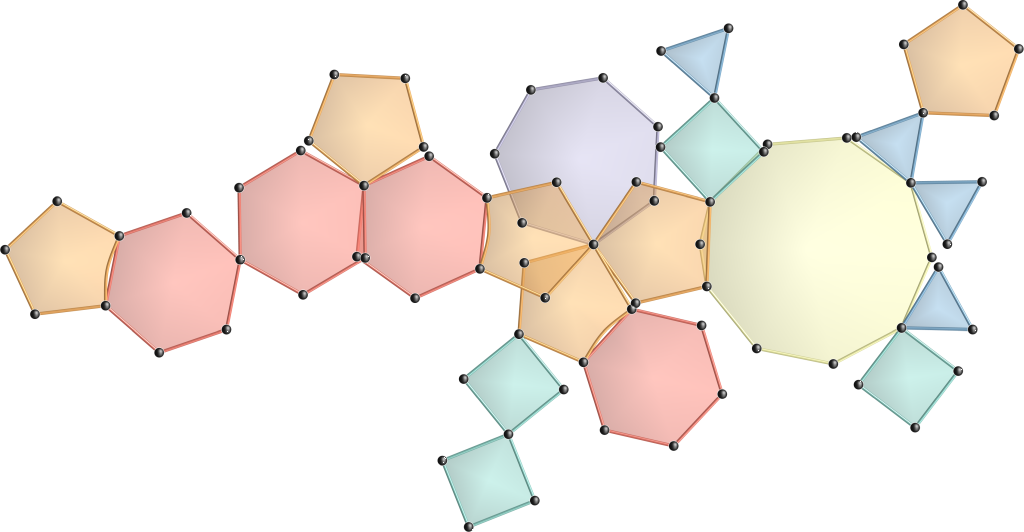}}\hspace{0.05in}
    \subfloat[][EulerView]{\includegraphics[height=1.0in]{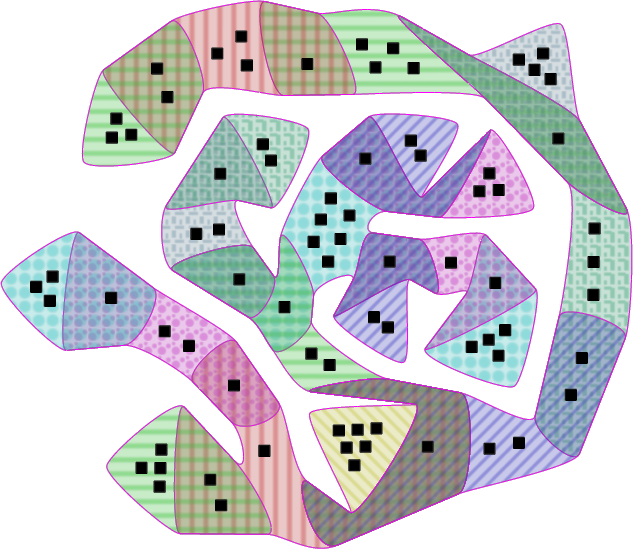}}\hspace{0.05in}
    \subfloat[][HyperVis]{\includegraphics[height=1.0in]{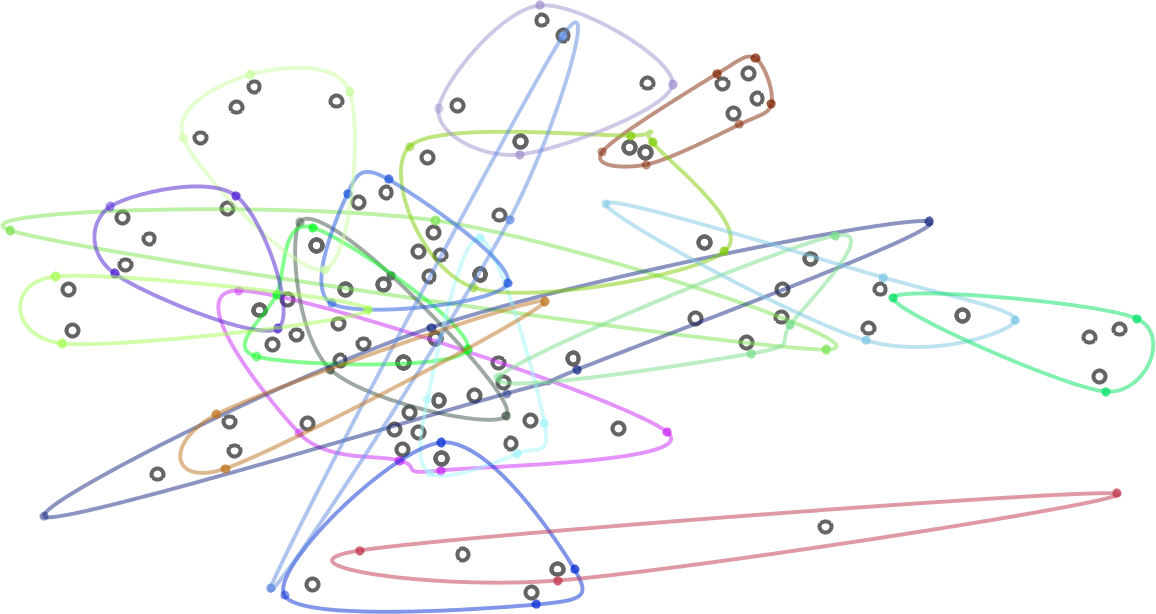}}\hspace{0.05in}
    \subfloat[][Zykov]{\includegraphics[height=1.0in]{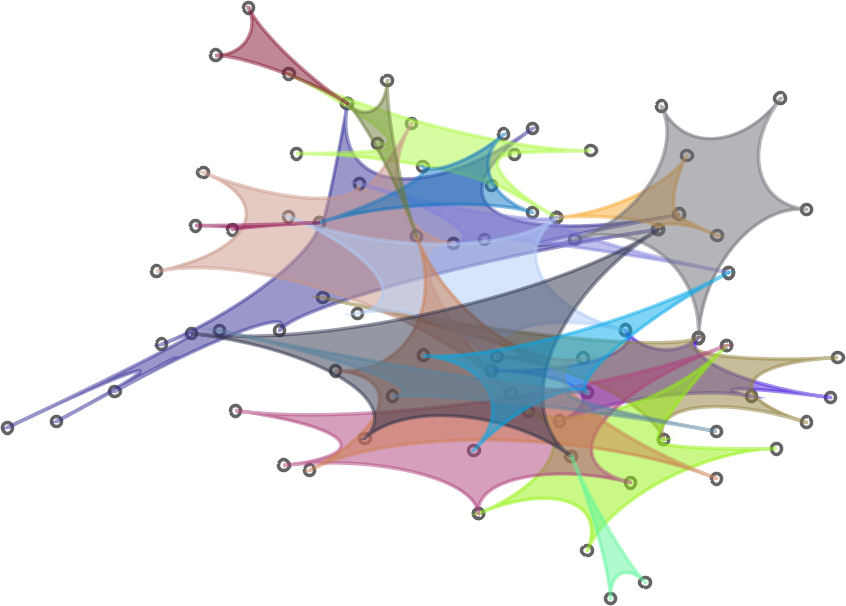}}
  \caption{Visualization results used for question Q2 in our user study with a dataset that contains $20$ hyperedges and $76$ vertices.}\label{fig:survey_images}
\end{figure*}

\section{Performance and Evaluation}
\label{sec:performance}

Our optimization framework requires repeated evaluation of the objective function given the current layout configuration (after starrization). Recall that $|V|$ and $|R|$ are the numbers of vertices and hyperedges in the data, respectively. The computation of the PR and PA energies have a complexity of $O(|V|+|R|)$. On the other hand, the computation of the PS and PI energies is more computationally costly as it requires processing each pair of hyperedges in the data. The complexity for the PS and PI energies is $O((|V|+|R|)^2)$. Similarly, each starrization operation has a complexity of $O((|V|+|R|)^2)$. This means that each evaluation of the objective function is of the complexity $O((|V|+|R|)^2)$. During each line search, there are a constant number of objective function evaluations. On the other hand, each sequence of pair swaps incur $O((|V|+|R|)^2)$ number of objective function evaluations, one per a potential pair swap. Thus, the complexity of each such sequence is $O((|V|+|R|)^4)$. Since the number of times of consecutive line searches and pair swaps depends on both the sizes of the data and the convergence criteria, our optimization framework has an overall complexity of $\omega((|V|+|R|)^4)$.

We have tested our optimization framework on eight datasets. Six of the datasets were collected from the DBLP database~\cite{author_data} with different search criteria and two were from an infectious disease dataset~\cite{isella2011s}. The smallest dataset has $76$ entities and $26$ relationships, while the largest dataset has $527$ entities and $232$ relationships. The time to perform the automatic optimization (either primal or dual) ranges from $0.25$ seconds for the smallest dataset to $181.69$ seconds for the largest dataset. When performing joint optimization, the time ranges from $1.37$ seconds for the smallest data to $268.82$ seconds for the largest data. The timing results were taken from a computer with an Intel(R) Xeon(R) E-$2124$G ~CPU @ $3.4$ GHz and $64$ GB RAM.

We conducted a user study with $21$ participants including five undergraduate students and $16$ graduate students to understand how the polygon-based layout compares to existing subset-based hypergraph visualization techniques and whether the incorporation of the dual view helps with data analysis.

Due to the ongoing COVID-19 pandemic and the resulting university closure, the user study was conducted remotely. The participants of our study consisted of $19$ students in Computer Science, $1$ student in Physics, and $1$ student in Mechanical Engineering. To our knowledge, the participants were not familiar with hypergraphs and their visualization. They were given five questions in the form of an online survey. Their answers and the time it took a participant to answer each question were recorded. The questions were as follows:

\begin{enumerate}
  \item[Q1] How many authors are part of the paper with the most co-authors?
  \item[Q2] Which layout helps you most effectively determine whether the two papers with the most co-authors share an author?\label{Q2}
  \item[Q3] Which layout helps you most effectively find the paper with the greatest number of authors?\label{Q3}
  \item[Q4] How many papers does the most productive author have?\label{Q4}
  \item[Q5] How many authors does the most authored paper of the most productive author have?\label{Q5}
\end{enumerate}

The first three questions were designed to compare our technique with three recent hypergraph visualization techniques that are region-based and whose software was available: EulerView~\cite{simonetto2009fully}, HyperVis~\cite{Arafat:17}, and the Zykov representation~\cite{zykov1974hypergraphs, ouvrard2020hypergraphs} with the layout generation method based on wheel graph introduced in~\cite{Arafat:17}. The results of these techniques were generated by using the released codes of these methods so that the colors and layouts were consistent with the respective published work. To provide some context to the questions, we used the data based on author-paper interpretation. That is, each polygon was a paper and the vertices of the polygon represented its authors. In our survey, the colors of the polygons were based on qualitative color schemes of ColorBrewer~\cite{harrower2003colorbrewer}, which are designed for categorical attributes. Thus, the participants still needed to extract the cardinality of a polygon based on the number of edges in the polygon.

Question~Q1 consisted of four individual questions. For each sub-question, the participants were given one of the four region-based hypergraph layouts: (a) EulerView, (b) HyperVis, (c) Zykov, and (d) N-ary (ours). Note that the layouts were based on the same dataset created synthetically, with $12$ papers (polygons) over $36$ authors (vertices). The order of the sub-questions was randomized and differed from user to user.
As shown in Figure~\ref{fig:survey_results} (a), the accuracy rate was the highest for our method. While the accuracy rate for Zykov's approach was also high, the average time to answer the question was $60$ seconds versus $40$ seconds for our approach. The standard errors of the data are also shown, which are comparable for the timing of all four techniques. On the other hand, the standard error is zero for our technique, indicating that all participants answered Q1 correctly.

Questions~Q2 and~Q3 were designed to understand the effectiveness of the four hypergraph visualization methods in conveying the relationships among hyperedges ($N$-ary relationships) in the data and the distribution of the cardinalities of $N$-ary relationships, respectively. Question~Q2 used a synthetic dataset with $20$ papers from $76$ authors, which was created from a larger actual dataset.
Question~Q3 used a more complex dataset with $95$ papers from $219$ authors which was also created from a larger actual dataset.
For both questions, the users were given all four layouts simultaneously, such as the one shown in Figure~\ref{fig:survey_images} for Question~Q2. As shown in Figure~\ref{fig:survey_results} (b-c), for both questions $95\%$ of the users favored our visualization over the other techniques.

Overall, our user study suggests that our visualization technique leads to higher accuracy and less time to finish a task than the other techniques. In addition, the participants appeared to prefer the polygon-based visualization over the other layouts.

The last two questions were designed to understand the potential benefits of including a dual view in the visualization. In our context, a polygon in the dual view is an author, and the vertices in the polygon are papers of the author.

Question~Q4 consisted of two sub-questions. In the first sub-question, the participants were given the primal view, while in the second they were given the dual view. To avoid bias, the users were not notified that these images were based on the same dataset.  As shown in Figure~\ref{fig:survey_results} (d), while achieving the same accuracy of $90\%$, the dual view required an average of $30$ seconds to complete the task while the primal view required an average of $75$ seconds to complete the same task. The dataset used in this question was the same as that for Question~Q3.

Question~Q5 consisted of three sub-questions: (1) the primal view only, (2) the dual view only, and (3) simultaneous display of both the primal and the dual that were jointly optimized. The dataset for this question was the same as that for Question~Q2. Again, the users were not notified that these visualizations were based on the same dataset. As shown in Figure~\ref{fig:survey_results} (d), for this question the dual view and the combined view led to the same accuracy ($81\%$) which is better than the primal view ($67\%$). With the combined view the participants needed an average of $40$ seconds to finish the task while with the dual view the participants needed on average $46$ seconds.

These results indicate potential benefits of the primal-dual view over using only the primal view. However, the pandemic-related university closure placed a number of constraints on our user study, such as the relatively small number of participants, the lack of control over the devices used in the survey (and thus the display sizes), and attention spans of the participants that affected our decision on the length of our survey to avoid incomplete answers. Consequently, we consider the findings of our user study {\em preliminary}. A more thorough, in-person user study with a controlled environment, is needed to validate or invalidate our preliminary findings.

\begin{figure*}[ht]
  \centering
\subfloat[][jointly optimized primal view]{\includegraphics[width=0.45\linewidth]{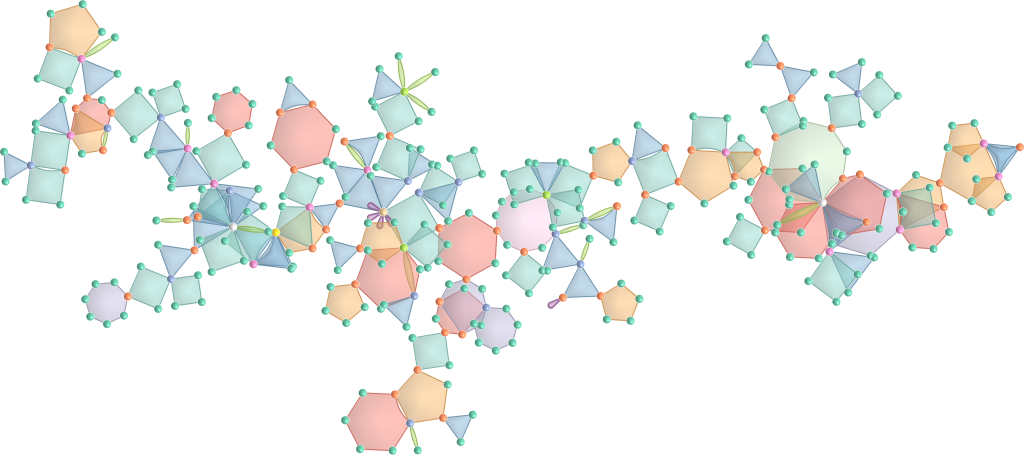}}\hspace{0.05in}
\subfloat[][jointly optimized dual view]{\includegraphics[width=0.45\linewidth]{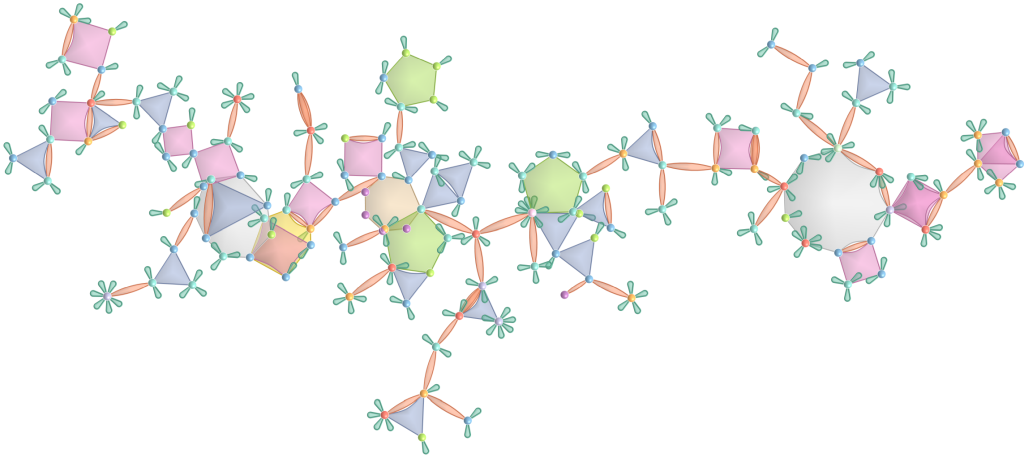}}
\hspace{0.2in}
  \caption{Paper and authorship data from the online database DBLP~\cite{author_data} for publications from 2013 to 2015 in IEEE Transactions on Pattern Analysis and Machine Intelligence. Each $N$-ary relationship is either a paper with $N$ authors (left: the primal view) or an author with $N$ papers (right: the dual view).}
  \label{fig:TRAMI}
\end{figure*}

\section{Case Studies}
\label{sec:applications}

We have applied our visualization technique to two applications: an author collaboration network and an infectious social contact network\cite{isella2011s}. %We discuss the author collaboration network in this section and the infectious social contact network in the Appendix~(Section~\ref{sec:friendship}).

\subsection{Authorship Collaboration Network}
\label{sec:author}

In Figure~\ref{fig:TRAMI}, we show the largest connected subset of publications for IEEE Transactions on Pattern Analysis and Machine Intelligence from $2013$ to $2015$. We make a number of observations.

From the primal view ((a): polygons$=$papers), we can see that there is a variety in the number of co-authors in a paper, ranging from two-author papers to eight-author papers. Combined with the lack of single-author papers (monogons), this highlights the collaborative nature of the field. In addition, it seems that the majority of the papers have three or four authors, indicating that this is the range of the number of people in a team that balances between productivity and management.

From the dual view ((b): polygons$=$authors), we observe that there are many authors with only one publication. We speculate that they were the students on the team, who after graduation, left the academic world of publications. On the other hand, polygons adjacent to many monogons are likely senior researchers and Ph.D. advisors. There are two highly productive authors (indicated by the grey polygons), who appear to be well-connected in the network but at a large distance. This can be perceived as the research areas that the two authors have worked on are relatively unrelated.

The {\em Erd\H{o}s number} measures the collaborative distance between a researcher and the Hungarian mathematician Paul Erd\H{o}s. In the research community in our dataset, we can similarly define such a distance between any researcher to the most productive author in the data. This is achieved by first identifying the largest polygon in the dual view (b), then finding the corresponding vertex in the primal view, and finally measuring the polygon distance between this vertex and the vertex representing the researcher whose Erd\H{o}s number is being computed.

Such insight is facilitated by the layouts generated by our automatic framework, and the primal-dual approach.

\begin{figure}[!hb]
\centering%
\subfloat[][primal view]{\includegraphics[width=0.45\linewidth]{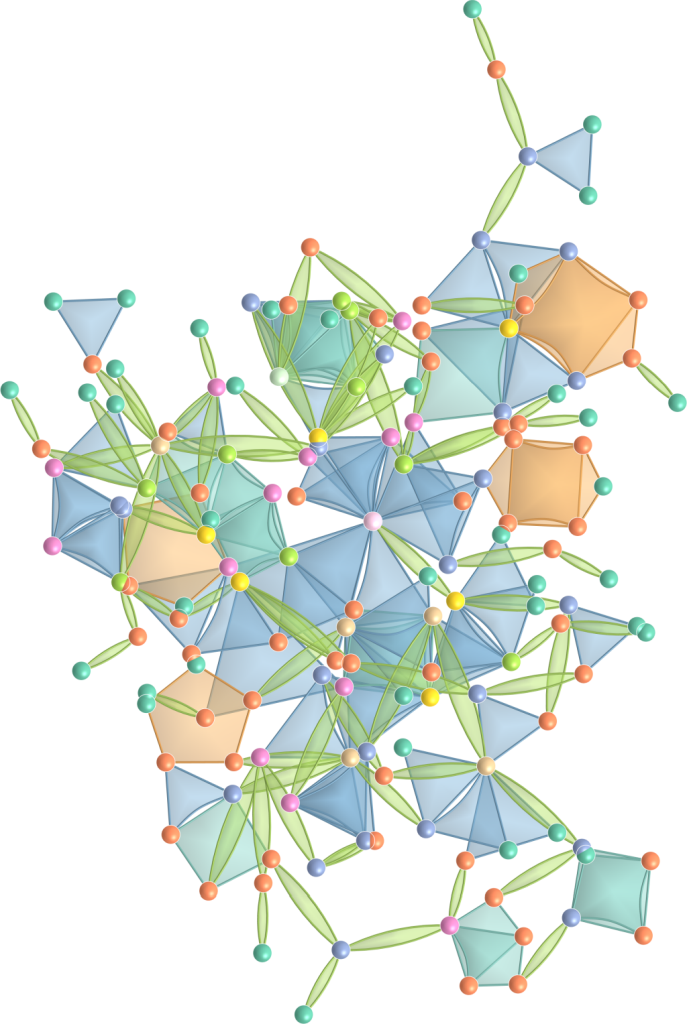}}
\subfloat[][dual view]{\includegraphics[width=0.45\linewidth]{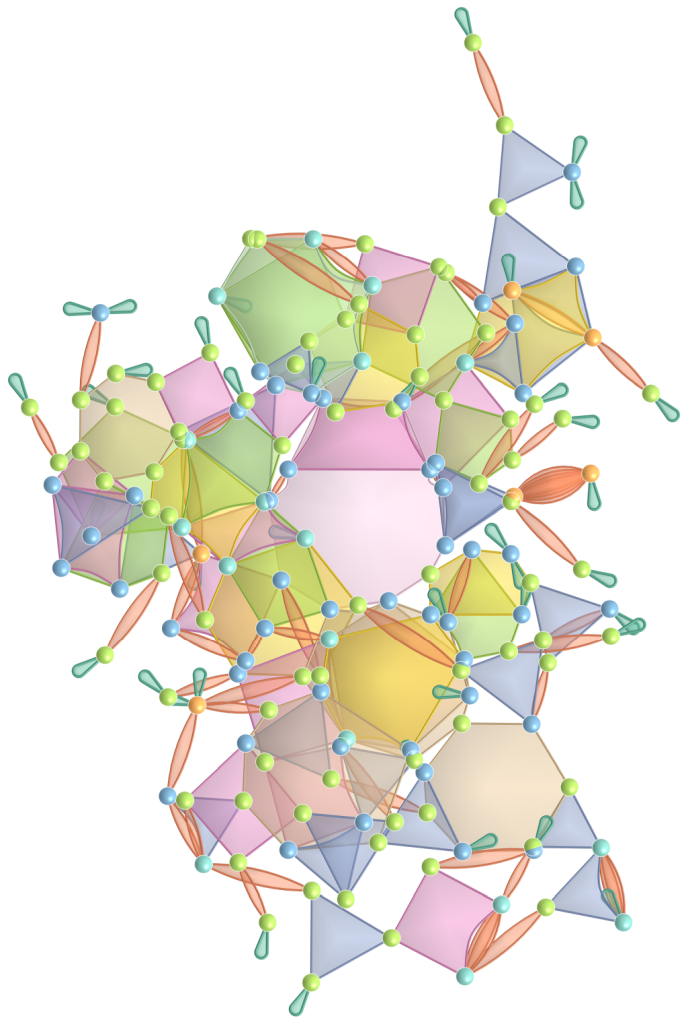}}\\
\subfloat[][primal view]{\includegraphics[width=0.45\linewidth]{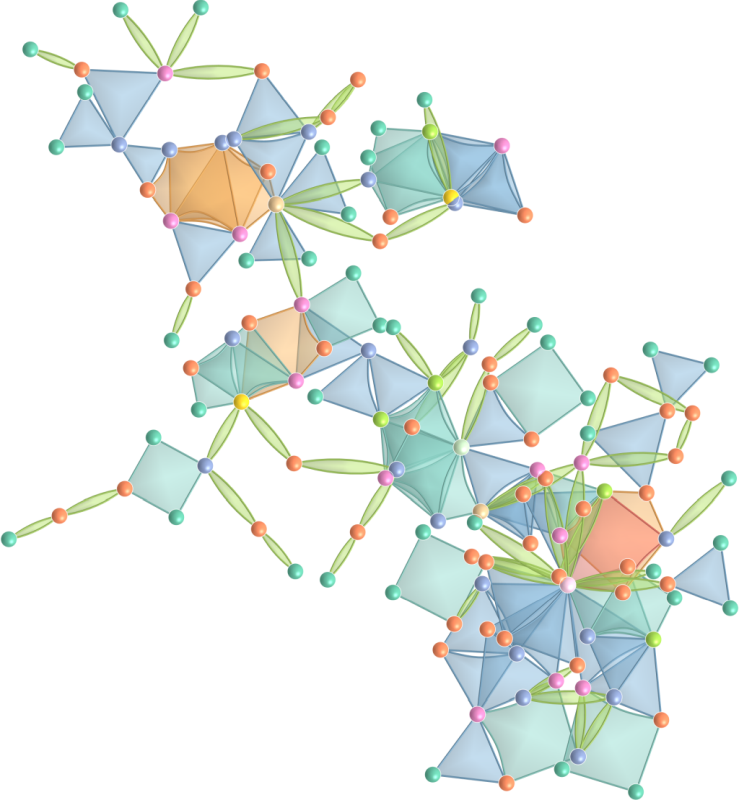}}
\subfloat[][dual view]{\includegraphics[width=0.45\linewidth]{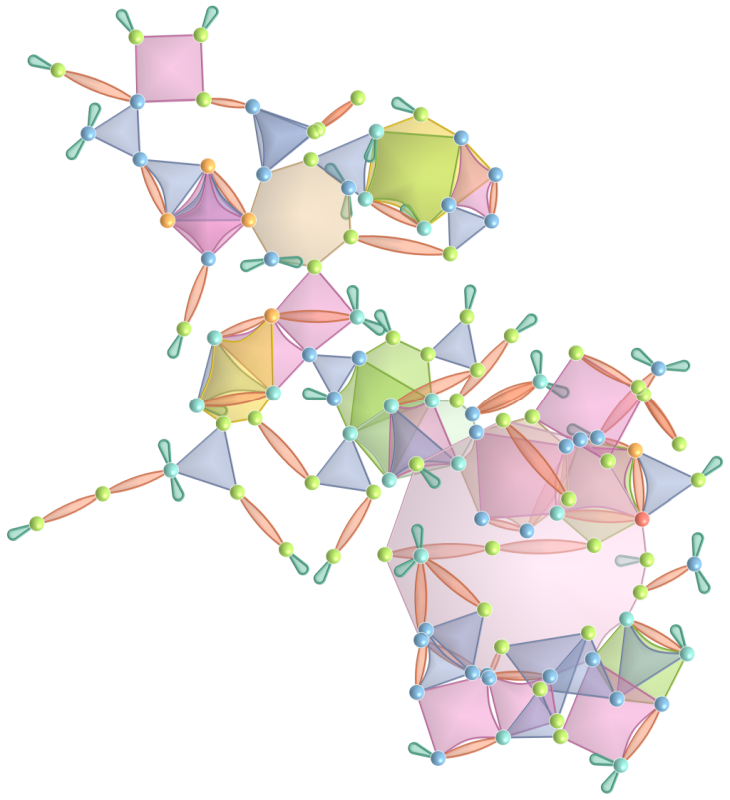}}\\
\caption[]{The visualization of a social contact pattern data~\cite{isella2011s} for May $10$, $2009$ (top) and June $20$, 2009 (bottom). The left images show the primal views and the right images show the dual views. Notice the difference in the patterns of the two days (top vs. bottom). }
\label{fig:infectious_pattern}
\end{figure}

\subsection{Social Contact Patterns}
\label{sec:friendship}

We also apply our framework to a social contact pattern dataset that aims to track the spread of infectious diseases. Isella et al.\cite{isella2011s} conduct an experiment by tracking visitors to a science gallery. The visitors are asked to wear an electronic badge that detects face-to-face contact. Unfortunately, the published data contains only the time durations of the contacts but not the physical distances of the contacts.

In the primal view, each visitor is an entity (vertex). A relationship (polygon) involving $N$ visitors implies that a {\em maximal} set of visitors who, as pairs, have spent more than $40$ seconds in a deemed close-enough distance in this study. We select two days (Sunday, May $10$, 2009 and Saturday, June $20$, 2009) from this dataset as our test cases, and visualize the data for both days in Figure~\ref{fig:infectious_pattern}.

From the primal views ((a) and (c)), we observe that there were more visitors on May $10$ than on June $20$. We speculate that this is partially due to the fact that more visitors visit the galley on Sundays than on Saturdays. It is also possible that more people choose to be outdoors in June when it is more likely to be sunny than in May when it can be rainy.

In addition, a large polygon corresponds to a group of people who had pairwise close contact. While they could be in close contact at different places and times, the fact that many visitors do not stay in the galley for a long period of time indicates that the group either knew each other (e.g. a family or a group of friends) or corresponds to a time of the day during which the members of the group visited the galley simultaneously (though uncoordinated).

The relatively many overlapping polygons are an indication that the groups of visitors may belong to a larger group. Recall that whether two visitors are considered in closed contact depends on whether they were in a close distance for over $40$ seconds. Both the distance threshold and the time threshold ($40$ seconds) were arbitrarily chosen. Varying the values of the thresholds and observing their impacts on the data can lead to additional insight for the application.

Finally, the large polygons in the dual views ((b) and (d)) correspond to people who had been in contact with the largest number of other visitors. This can be caused by them being in the galley longer than other visitors, such as the employees of the gallery or the tour guides working there. Regardless, should an infectious disease break out, such people can contribute to the faster spread of the disease. Tracing their activities before and during the breakout can be more urgent than tracking the visitors represented by the smaller polygons.

The above observations and hypotheses are enabled by our automatic layout optimization technique and the primal-dual approach.

\section{Conclusion and Future Work}
\label{sec:conclusion}

The main contribution of our work is the introduction of an automatic polygon layout optimization framework that enhances the quality of the layout for hypergraphs. At the core of our technique is a set of design principles for polygon layout that we have identified and the objective function based on these principles. To our knowledge, it is the first time that the order of vertices in a polygon is explicitly addressed during layout generation (the pair swap operation and the polygon intersection energy). To avoid self-overlaps and flips in the layout, we develop a procedure called starrization which guarantees that the layout is free of such artifacts. We also handle datasets with monogons.

Recognizing the duality between entities (vertices) and relationships (hyperedges), our system enables simultaneous display of both the primal view and the dual view. To correlate the two layouts (primal and dual), we enable an automatic joint layout optimization framework based on an augmented energy term that encourages the spatial correlations between the two views.

Through a user study, we show that the polygon-based layouts generated by our automatic framework compare favorably over a number of recent subset-based hypergraph approaches for a number of tasks. In addition, the user study confirms the benefits of the primal-dual visualization framework and the joint optimization.

There are limitations to both the polygon metaphor and our optimization approach. As our optimization algorithm has a complexity of $\omega(N^4)$ where $N$ is the sum of the number of vertices and the number of hyperedges, it can be difficult to be scaled up to much larger datasets. We plan to explore hierarchical optimization and visualization to alleviate this problem. In addition, we will investigate the use of the GPUs as the evaluation of the objective function is highly parallelizable.  Another issue with our optimization is that our polygon area (PA) energy and polygon separation (PS) energy are formulated assuming the underlying polygon has a low polygon regularity (PR) energy, i.e. being nearly regular. Our choice of such formulations is motivated by a number of factors such as reducing the computational cost. For example, finding the exact distance between two polygons not sharing a vertex is a classical computational geometry problem. Employing a formulation that requires exact computation can further increase the computational complexity of our optimization framework. However, when the polygons are not close to being regular, our current formulations of the PA and PS energies are no longer as effective. We will investigate other formulations that are less dependent on the regularity of the polygons.

The polygon metaphor also has its limitations. It performs best when the underlying hypergraph has approximately a tree-like structure. When there are an excessive number of polygons adjacent to a vertex, overlaps among these polygons are unavoidable. Similarly, a cluster of polygons can have unavoidable overlaps (e.g. Figure~\ref{fig:infectious_pattern}). When these types of overlaps occur, which can become more prominent as the tree-like structure in the data disappears, we observe that the effectiveness of the polygon metaphor decreases. Tasks such as recognizing a hyperedge and its cardinality, deciding whether a vertex belongs to a hyperedge, and perceiving whether two hyperedges intersect become more difficult. We plan to explore a multi-scale representation of hypergraphs to address this challenge. In addition, we plan to investigate the adaptation of graph sparification~\cite{Lai:2020} to hypergraph sparcification, in which less important data is filtered out from the visualization.

For visualization, we will explore better layouts to reduce the amount of unused space. Incorporating label placement for vertices and hyperedges in both the primal view and the dual view can strengthen the link between the two views. This is a potentially fruitful research direction. Addressing the uncertainty in the data is another future research avenue.

\acknowledgments{The authors wish to thank our anonymous reviewers for their constructive feedback. We appreciate the help from Avery Stauber during video production. We also would like to thank Dr. Markus Wallinger and Dr. Danial Archambault for sharing the codes of EulerView. This work was supported in part by the NSF award (\# 1619383).}

\bibliographystyle{abbrv-doi}
\bibliography{nary}

	\clearpage
	\appendix

\section{Initial Layout and Its Impact on Final Layout}
\label{sec:appendix}

\begin{figure}[H]
     \centering
    $\begin{array}{@{\hspace{0.0in}}c@{\hspace{0.10in}}c}
    \includegraphics[height=2.5in]{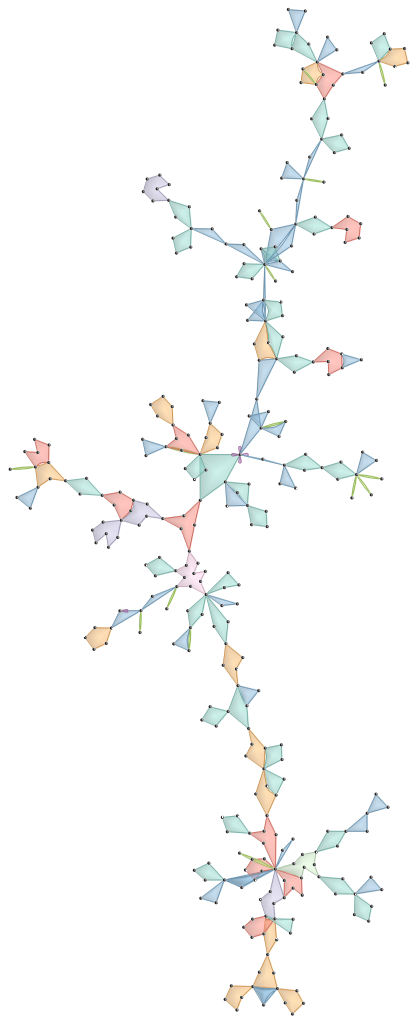}
 &\includegraphics[height=2.5in]{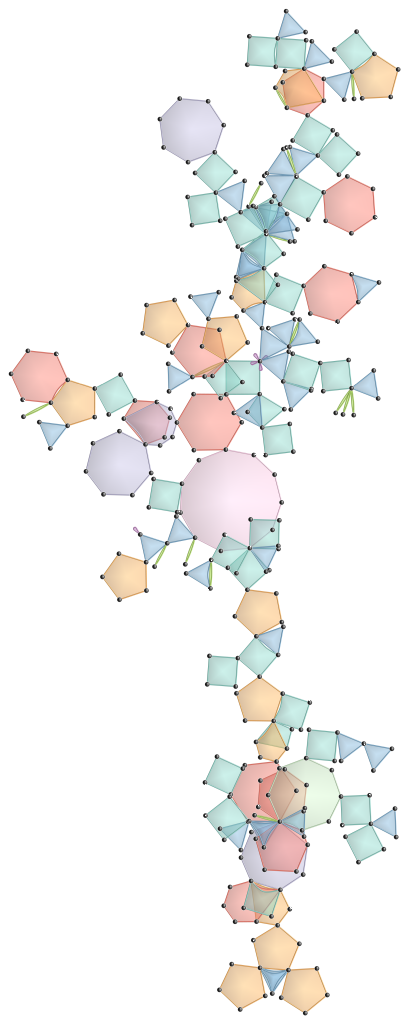} \\
(a) & (b)\\
  \end{array}$
     \caption{The optimized layout for a dataset (b) given an initial layout where the vertices are placed based on the force-directed layout algorithm of Hu~\cite{hu2005efficient} (a).
     }
\label{fig:initial_layout_FD}
     \centering
    $\begin{array}{@{\hspace{0.0in}}c@{\hspace{0.10in}}c}
    \includegraphics[width=1.7in]{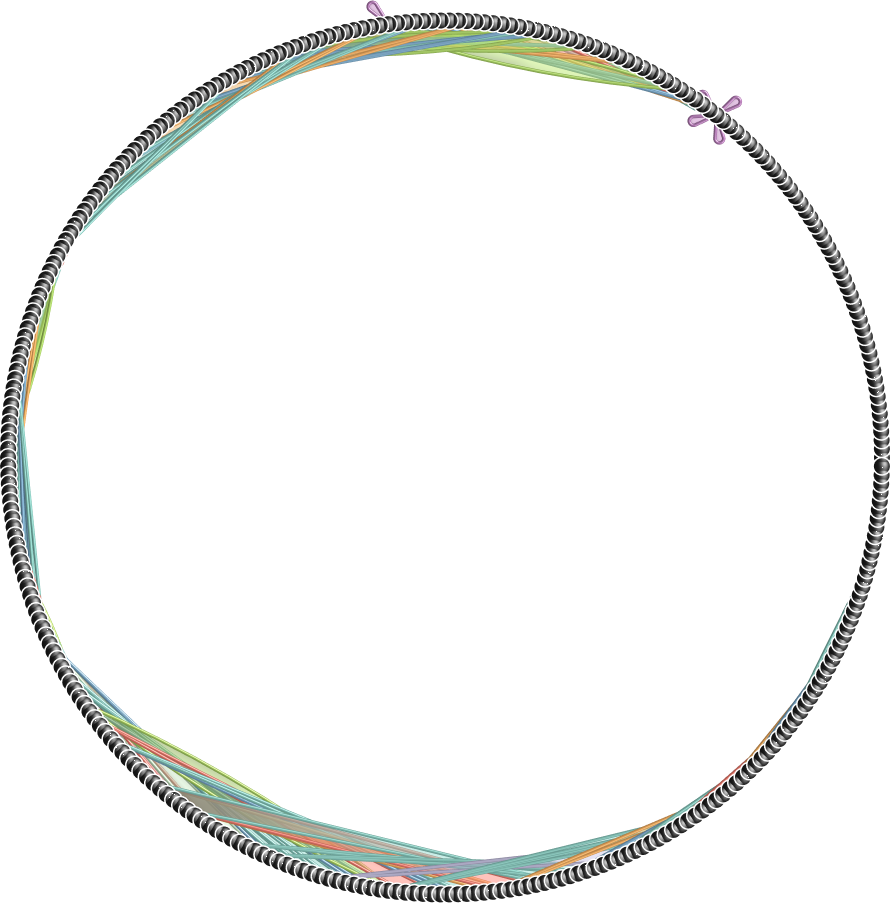}
 &\includegraphics[width=1.7in]{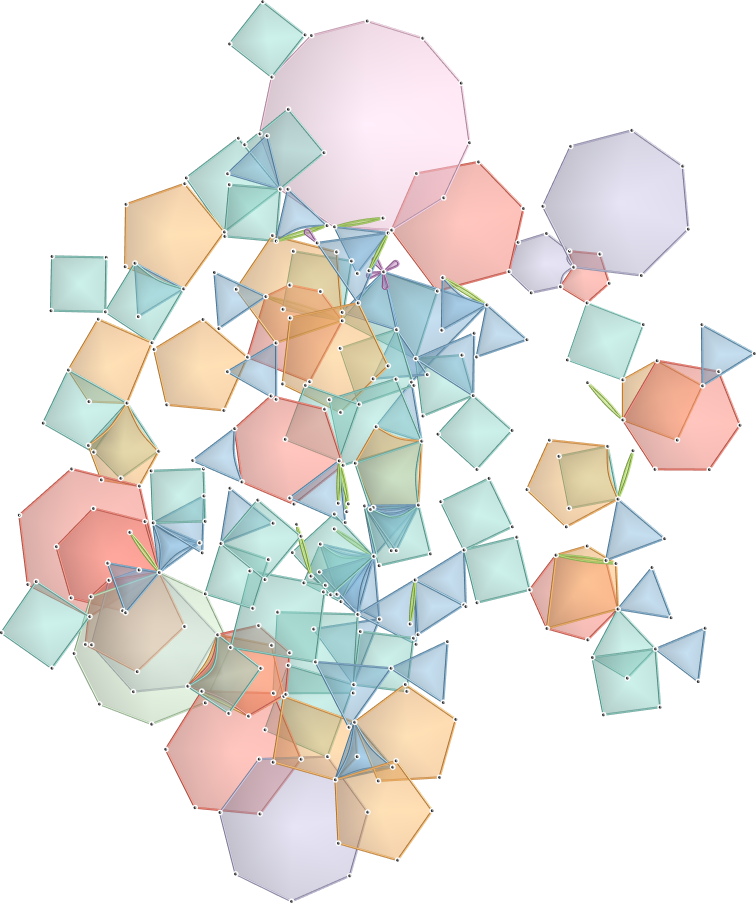} \\
(a) & (b)\\
  \end{array}$
     \caption{The optimized layout for a dataset (b) given an initial layout where the vertices are placed on a circle (a).
     }
\label{fig:initial_layout_circle}
     \centering
    $\begin{array}{@{\hspace{0.0in}}c@{\hspace{0.10in}}c}
    \includegraphics[width=1.7in]{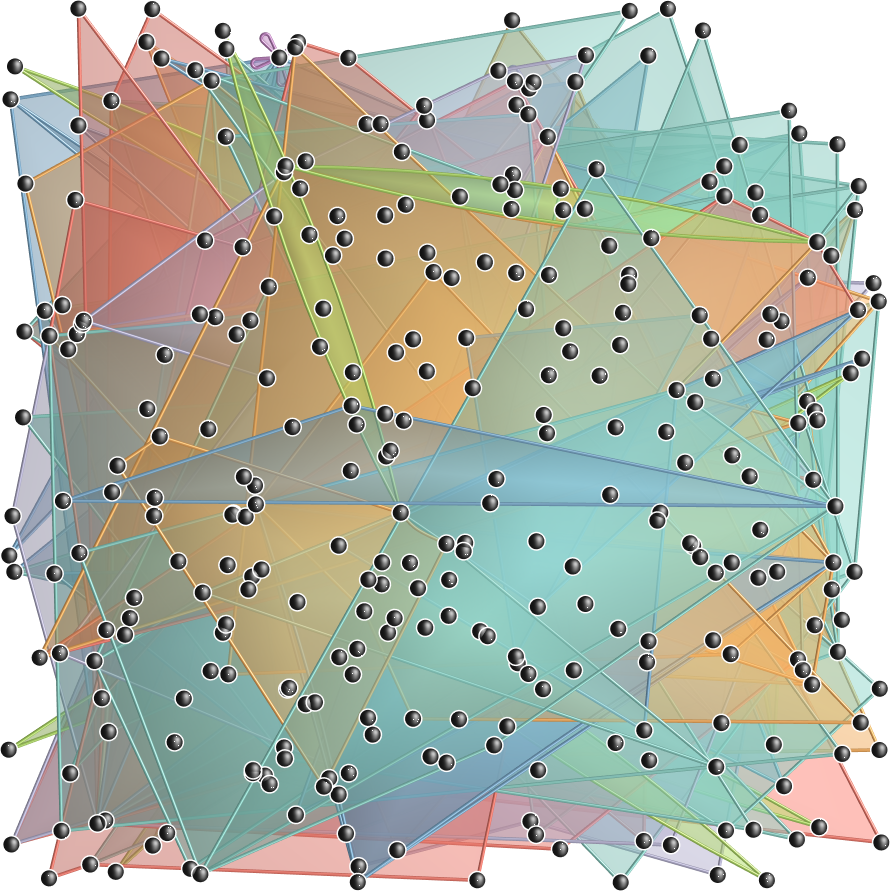}
 &\includegraphics[width=1.7in]{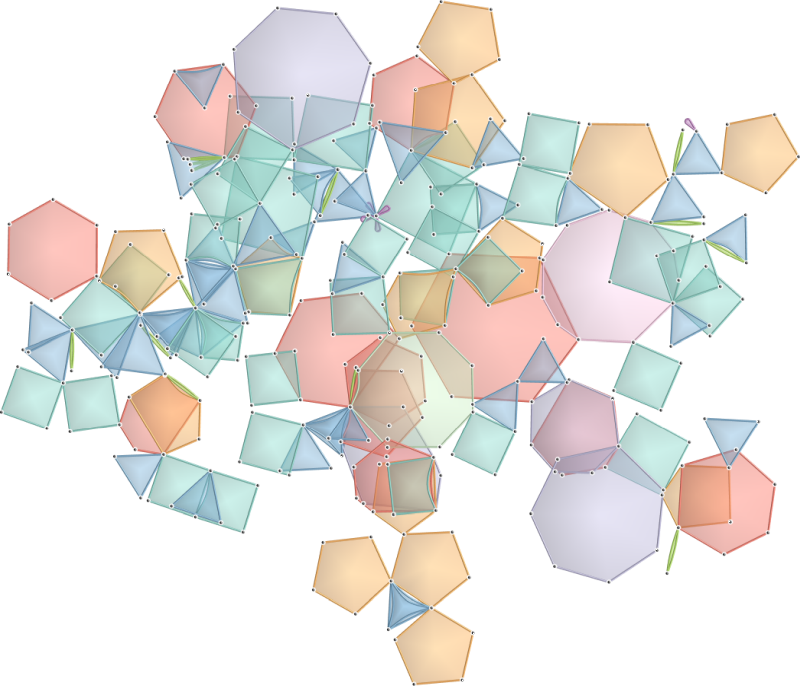} \\
(a) & (b)\\
  \end{array}$
     \caption{The optimized layout for a dataset (b) given an initial layout where the vertices are randomly placed (a).
     }
\label{fig:initial_layout_random}
\end{figure}

Different initial layouts of the same dataset can lead to different final optimized layouts: force-directed~\cite{hu2005efficient} (Figure~\ref{fig:initial_layout_FD}), circular (Figure~\ref{fig:initial_layout_circle}), and random (Figure~\ref{fig:initial_layout_random}). Our system provides all three initial layout schemes for the user. The results in the paper were based on the force-directed layout similar to Figure~\ref{fig:initial_layout_FD}.

\section{Time Comparison between Manual Layout Design and Automatic Optimization}
\label{sec:manual_and_opt}

Table~\ref{tbl:manual_and_opt} compares the time of manually designing a polygon layout to that of using our automatic optimization-based algorithm. Note that for all three test datasets, the automatic algorithm is about three magnitudes faster than manual design. Both the manual design and the automatic algorithm start with the same initial layout, which, for the three test datasets, are based on the force-directed method~\cite{hu2005efficient}.

\begin{table}[!h]
\centering
\begin{tabular}{|l||l|l|l|l|}
\hline
  Data    & No.     & No.        & Design time & Optimization         \\
      & Vertices     & Hyperedges         &(seconds) & time (seconds)        \\
      \hline\hline
No. 1   & $81$ & $22$  & $453$    & $0.670$  \\ \hline
No. 2   & $72$ & $22$  & $304$    & $0.737$   \\ \hline
No. 3   & $43$ & $14$ & $194$ & $0.268$  \\ \hline
\end{tabular}
\caption{This table compares the times of manually designing a polygon layout for three hypergraphs to those of using our automatic optimization framework.  }
\label{tbl:manual_and_opt}
\end{table}

\end{document}